\documentclass[%
 reprint,
superscriptaddress,
showkeys,
 amsmath,amssymb,
 aps,
 pre
]{revtex4-1}

\usepackage{graphicx}
\usepackage{dcolumn}
\usepackage{bm}
\usepackage{xcolor} 
\usepackage[normalem]{ulem} 

\newcommand{\md}{\mathrm d}
\newcommand{\Nbath}{N_{\rm tot}}
\newcommand{\Ntot}{N_{\rm tot}}
\newcommand{\Nub}{N_{\rm UB}}
\newcommand{\Nb}{N_{\rm B}}
\newcommand{\muI}{\mu_{\rm i}}
\newcommand{\muF}{\mu_{\rm f}}

\newcommand{\SB}{\color{black}}
\newcommand{\stkout}[1]{\ifmmode\text{\SB \sout{\ensuremath{#1}}}\else{\SB\sout{#1}}\fi}

\begin{document}

\title{Optimal control of protein copy number}

\author{Steven Blaber}
\affiliation{Dept.~of Physics, Simon Fraser University, Burnaby, British Columbia V5A 1S6, Canada}
\affiliation{Dept.~of Physics, University of Waterloo, Waterloo, Ontario N2L 3G1, Canada}
\author{David A.~Sivak}
\email{dsivak@sfu.ca}
\affiliation{Dept.~of Physics, Simon Fraser University, Burnaby, British Columbia V5A 1S6, Canada}

\date{\today}

\begin{abstract}
Cell-cell communication is often achieved by secreted signaling molecules that bind membrane-bound receptors. A common class of such receptors are G-protein coupled receptors, where extracellular binding induces changes on the membrane affinity near the receptor for certain cytosolic proteins, effectively altering their chemical potential. We analyze the minimum-dissipation schedules for dynamically changing chemical potential to induce steady-state changes in protein copy-number distributions, and illustrate with analytic solutions for linear chemical reaction networks. Protocols that change chemical potential on biologically relevant timescales are experimentally accessible using optogenetic manipulations, and our framework provides non-trivial predictions about functional dynamical cell-cell interactions.
\end{abstract}

\maketitle

\section{\label{sec:intro}Introduction}
Biochemical reaction networks play a central role in cellular response to external stimuli (such as cell-cell signaling), converting inter-cellular signals into a driven chemical response~\cite{signalling}.	
A prominent communication channel for chemical signals across the cell membrane is the G-protein coupled receptors (GPCRs). An agonist ligand binds to the extracellular face of a GPCR and allosterically induces a conformational change on its intracellular face. This conformational change stimulates exchange of GDP for GTP on the $\alpha$ subunit of the intracellularly bound heterotrimeric G-protein, thereby reducing the binding affinity between the G-protein and the receptor. The G-protein unbinds from the GPCR and dissociates into separate $\alpha$ and $\beta\gamma$ subunits, which respectively diffuse away from the GPCR in the cytosol and in the membrane~\cite{MBOC}. The delocalization of the $\alpha$ and $\beta\gamma$ subunits from the GPCR increases their concentration in the cytosol and at other membrane locations, respectively, which elicits a series of reactions ultimately leading to the downstream cellular response~\cite{nobel2012}. 

In mammals, GPCRs mediate many physiological responses---to changes in concentrations of peptides, hormones, lipids, neurotransmitters, ions, odorants, tastants, and light. Since $\sim$1000 human genes code for GPCRs~\cite{nobel1,nobel2}, we predict that a more energetically efficient signaling process through a GPCR (all else being equal) would provide a selective advantage, such that evolved signaling pathways could be expected to exhibit impressive efficiency. While energetic efficiency is surely only one of many criteria that influence natural selection, presumably greater efficiency provides an advantage when holding other criteria constant~\cite{Hasenstaub:2010eh}.

One ultimate effect of agonist ligand binding extracellularly to the GPCR is to decrease the integral membrane protein GPCR's intracellular binding affinity for the G-protein, and thus increase the concentration of G-protein $\alpha$ subunit in the cytosol. Thus the GPCR signaling process can be modeled as changes in the chemical potential difference between unbound G-protein and G-protein bound at the cell membrane
(from hereon simply referred to as the chemical potential), ultimately driving changes in G-protein concentration in the cytosol. The chemical potential is externally controlled by modulating the number of activated GPCRs through, for example, changing extracellular concentrations of agonist ligand.

For given desired equilibrium endpoints of chemical potential, any protocol (schedule of changing chemical potential) that proceeds quasi-statically (at negligible speed) requires the same input energy in the form of chemical potential work, an amount equal to the free energy change between the equilibrium ensembles at the two endpoint chemical potentials. For protocols that proceed at a finite velocity, different protocols differ in their energetic costs, and hence in the required number of signaling molecules the signaling cell must secrete.

Here we develop theory describing how a cell can achieve a given dynamic signaling outcome at minimal energetic cost. This can be formalized in the language of a previously developed theoretical framework in nonequilibrium statistical mechanics, that of finding a protocol that minimizes the excess work associated with finite-time changes in a control parameter~\cite{molecularmachines}. Starting from a theoretical framework developed in \cite{OptimalPaths} to approximate the thermodynamic cost (excess work) of rapid changes in an arbitrary control parameter, we extend the formalism to address changes in chemical potential, and derive protocols that minimize the required work.

We find that near equilibrium, the excess work is determined by the auto-covariance of the protein copy number. For the special case of linear-order chemical reactions, we derive analytic forms of the generalized friction tensor, and the required work for both designed and naive (constant-velocity) protocols. We illustrate these results in simple chemical reaction schemes: an open system exchanging molecules with a molecular reservoir, and a closed system with fixed total copy number.

\section{\label{sec:theoryReview}Theoretical review}

We first present a review of minimum-dissipation nonequilibrium control in the linear-response framework. {\SB The average excess power (above the average power if the system were equilibrated throughout the driving protocol) exerted by an external agent changing control parameters $\lambda$ that are coupled to the system in the canonical ensemble is~\cite{OptimalPaths}
\begin{align}
\md_t W_{\rm ex} = -\langle \delta f_{j} \rangle_{\Lambda} \md_t \lambda_{j} \ .
\label{General exact power}
\end{align} 
Here $\md_t$ denotes the time derivative, $\beta \equiv (k_{\rm B}T)^{-1}$ is inverse temperature, $f_{j} \equiv -\partial_{\lambda_{j}} U$ is the force conjugate to the $j$th control parameter, and $\delta f_{j}(t) \equiv f_{j}(t) - \langle f_{j}\rangle_{\lambda}$ are the equilibrium fluctuations. $\langle \cdots \rangle_{\lambda}$ indicates an equilibrium average for fixed $\lambda$, and $\langle \cdots \rangle_{\Lambda}$ a non-equilibrium average during the control parameter protocol $\Lambda$. Throughout, we adopt the Einstein summation convention of implied summation over all repeated indices.}
Applying linear-response theory~\cite{OptimalPaths} gives a near-equilibrium expression,
\begin{align}
\label{LR excess power}
\md_t W_{\rm ex}(t) \approx \md_t \lambda_{j} \ \zeta_{j \ell}[\lambda(t)] \ \md_t \lambda_{\ell} \ ,
\end{align}
in terms of the generalized friction tensor
\begin{align}
\label{Zeta}
\zeta_{j \ell}(\lambda) \equiv \beta \int_0^{\infty} \md t \, \langle \delta f_{j}(t) \delta f_{\ell}(0)\rangle_{\lambda} \ {\SB,} 
\end{align}
{\SB with $\langle \delta f_{j}(t) \delta f_{\ell}(0)\rangle_{\lambda}$ the force covariance.}
	
The generalized friction tensor $\zeta_{j \ell}$ is the Hadamard product $\beta \langle\delta f_{j} \delta f_{\ell}\rangle_{\lambda} \circ \tau_{j \ell}$ of the conjugate force covariance (the force fluctuations) and the integral relaxation time
\begin{align}
\label{relax1}
\tau_{j \ell} \equiv \int_0^{\infty} \md t \frac{\langle \delta f_{j}(t) \delta f_{\ell}(0)\rangle_{\lambda}}{\langle \delta f_{j} \delta f_{\ell}\rangle_{\lambda}} \ ,
\end{align}
the characteristic time it takes for these fluctuations to die out.
	
The generalized friction tensor reflects the increased energy cost associated with rapid driving through control parameter space.
Integrating the excess power~\eqref{LR excess power} over the control parameter protocol gives the mean excess work, 
\begin{align}
\label{excess work}
W_{\rm ex} = \int_0^{\Delta t} \md t \ \md_t W_{\rm ex}(t) \ ,
\end{align}
above and beyond the quasi-static work. 

Under the linear-response approximation, the excess work is minimized for a `designed' protocol with constant excess power~\cite{OptimalPaths}. For a single control parameter, this amounts to proceeding with a velocity $\md_t \lambda^{\rm des} \propto \zeta(\lambda)^{-1/2}$, which when normalized to complete the protocol in a fixed allotted time $\Delta t$, gives 
\begin{align}
\label{lambdaoptdot2}
\md_t \lambda^{\rm des} = \frac{\int_{\lambda_{\rm i}}^{\lambda_{\rm f}}\md\lambda' \sqrt{\zeta(\lambda')}}{\sqrt{\zeta(\lambda)} \Delta t} \ ,
\end{align}
for initial and final {\SB control parameters} $\lambda_{\rm i}$ and $\lambda_{\rm f}$, respectively. 
	
Thus for a fixed protocol time, work is minimized by driving the system (changing the control parameter) slowly in regions of high friction, and quickly in areas of low friction.
The ratio of excess works during the naive and designed protocols is~\cite{Workratio}
\begin{align}
\label{Ratio}
\frac{W_{\rm ex}^{\rm naive}}{W_{\rm ex}^{\rm des}} = \frac{\Delta \lambda \int_{\lambda_{\rm i}}^{\lambda_{\rm f}} \zeta(\lambda) \, \md\lambda}{\left[\int_{\lambda_{\rm i}}^{\lambda_{\rm f}} \sqrt{\zeta(\lambda)} \, \md\lambda\right]^2} \ .
\end{align}

\section{Driving chemical potential}
A system of $n$ different chemical species at thermal and chemical equilibrium with a single heat reservoir and multiple particle reservoirs at temperature $T$ and chemical potentials $\mu_{j}$, respectively, is described by the grand canonical ensemble (GCE) with free energy (\emph{grand potential})
\begin{align}
\label{Free energy}
\Phi_{\rm G} \equiv U - TS - \mu_{j} N_{j} \ ,
\end{align}
for system energy $U$ {\SB (as in \S\ref{sec:theoryReview})}, entropy $S$, and copy number $N_{j}$ of the $j$th chemical species. {\SB To extend Eq.~\eqref{General exact power} to the GCE we show in Appendix~\ref{Exact Work} that the appropriate conjugate force is $f_{j}=-\partial_{\lambda_{j}}\Phi_{\rm G}$.} In this study, the control parameters $\lambda_{{\SB j}}$ are chemical potentials $\mu_{{\SB j}}$, and hence the conjugate forces are the copy numbers, $f_{j} = -\partial_{\mu_{j}} \Phi_{\rm G} = N_{j}$.

This produces a friction tensor and excess work
\begin{subequations}
\begin{align}
\label{Zeta cov}
\zeta_{j \ell}(\mu) &= \beta \, \int_0^{\infty} \md t \langle \delta N_{j}(t) \delta N_{\ell}(0)\rangle_{\mu} \\
\label{Free energy friction}
&= \beta \langle \delta N_{j} \delta N_{\ell}\rangle_{\mu}\circ \tau_{j \ell}(\mu) \\
\label{Free Energy Work}
W_{\rm ex} &= \beta \int_0^{\Delta t} \md t \, \md_t \mu_{j} \langle \delta N_{j} \delta N_{\ell}\rangle_{\mu}\circ \tau_{j \ell}(\mu) \md_t \mu_{\ell} \ .
\end{align}
\end{subequations}
The total work during a chemical-potential protocol is the equilibrium free energy change, plus an additional contribution from the excess work. This extra cost is proportional to the relaxation time $\tau$ and equilibrium copy-number covariance $\langle \delta N_{j} \delta N_{\ell}\rangle_{\mu}$, so 
{\SB rapidly changing the chemical potential incurs greater energy cost (due to system resistance) in}
reaction systems subject to large and long-persisting fluctuations in protein copy number.
{\SB Such continuous changes of chemical potential are plausible in natural settings: the chemical potential is a function of the ligand-binding state of the collection of receptors, so for more than a few receptors, the chemical potential changes fairly smoothly even upon rather sudden extracellular concentration changes, as the receptors progressively bind ligand (or progressively unbind upon ligand depletion).}

\section{\label{Linear Markov chemical reaction networks}Linear Markov chemical reaction networks}
The dependence of the friction tensor $\zeta$ on control parameter $\mu$, and thus the solution for the designed protocol, is a function of the topology and kinetics of the chemical reaction network~\cite{Closedfirstorder,Firstorderreactions,Momentclosure}.
Here we model the stochastic behavior of chemical reaction systems assuming Markovian dynamics, where the future dynamics depends exclusively on the present state.

{\SB In general, the autocovariance for non-linear chemical reaction networks cannot be solved analytically. A conceptually simple alternative is to numerically calculate the autocovariance~\cite{Lucero:2019gd} using a stochastic simulation of the chemical reaction dynamics, such as the Gillespie algorithm~\cite{Gillespie}; however, direct simulations can be computationally intensive. An alternate approach is to find approximate solutions using moment-closure techniques~\cite{Momentclosure}. 
Briefly, the chemical master equation~\cite{Gardiner,vanKampen}
leads to coupled ordinary differential equations describing evolution of the moments of the probability distribution of chemical counts. Approximations, that permit expression of higher-order moments in terms of lower-order moments, lead to the dynamics of the entire probability distribution being described by a small number of moment-evolution equations, which can be solved to find the equilibrium autocovariance. \cite{Momentclosure} provides more detailed discussion. 

For linear-order chemical reactions, the autocovariance---and therefore the friction tensor---can be solved exactly~\cite{Closedfirstorder,Firstorderreactions}.}
A linear-order chemical reaction system with multiple chemical species (and fixed chemical potential) satisfies~\cite{Closedfirstorder}
\begin{align}
\label{mean evolution}
\md_t \overline{N_{j}(t)} = -K_{j \ell} \overline{N_{\ell}(t)} + k^{\rm s}_{j} \ ,
\end{align}
where $K \equiv K^{\rm d} - K^{\rm con}$, $K^{\rm d}$ is the diagonal matrix of degradation rates, $K^{\rm con}$ is the matrix of conversion reaction rates, 
{\SB and} 
$k^{\rm s}$ are the production rates from a constant source.
An overbar indicates an (in general out-of-equilibrium) average {\SB $\overline{N_{j}(t)} = \int \md N_{j} \, N_{j} \, p(N_{j},t|N_{j}(t_0),t_0)$, with $ p(N_{j},t|N_{j}(t_0),t_0)$ the conditional probability of finding $N_{j}$ molecules at time $t$, subject to the initial condition $N_{j}(t_0)$ at time $t_0$}.
For notational simplicity, in this section we suppress explicit dependence on $\mu$.

Equation~\eqref{mean evolution} has the general solution
\begin{align}
\label{Mean time evolution}
\overline{N_{j}(t)} = &\left[e^{-K t}\right]_{j \ell} N_{\ell}(0) \\
&+\left({\SB \delta_{jm} }-\left[e^{-K t}\right]_{j{\SB m} }\right) \int_0^{t}\md t'\left[e^{-K t'}\right]_{{\SB m}\ell}k^{\rm s}_{\ell} \ . \nonumber
\end{align}
Assuming $K$ is diagonalizable, then $e^{-K t} = V e^{-D{\SB t}}V^{-1}$, where $D$ is the diagonal eigenvalue matrix, and $V$ is the eigenvector matrix, whose rows are the corresponding eigenvectors of $K$. If $K$ is not diagonalizable, then other standard methods of computing the matrix exponential can be employed~\cite{matrixexponential,newmatrixexponential}.

For a linear Markov reaction network, the auto-covariance obeys a similar time evolution equation as the mean~\cite{Gardiner}:
\begin{align}
\label{cov evolution}
\md_t \langle \delta N_{j}(t) \delta N_{\ell}(0)\rangle = -K_{jm}\langle \delta N_{m}(t) \delta N_{\ell}(0)\rangle \ .
\end{align}
Assuming the system is initially at equilibrium, this has the solution
\begin{align}
\label{autocovariance}
\langle \delta N_{j}(t) \delta N_{\ell}(0)\rangle &= \left[e^{-K t}\right]_{jm}\langle \delta N_{m}\delta N_{\ell}\rangle \\
&= V_{j {\SB n}}{\SB e^{-\lambda_{n}t}}[V^{-1}]_{n {\SB m}}\langle \delta N_{{\SB m}}\delta N_{\ell}\rangle \ .
\end{align}

This produces a friction tensor 
\begin{align}
\zeta_{j \ell} 
\label{zeta}
&= \beta \int_0^{\infty}\md t \, V_{j {\SB n}}{\SB e^{-\lambda_{n}t}}[V^{-1}]_{n {\SB m}}\langle \delta N_{{\SB m}}\delta N_{\ell}\rangle \\
\label{zeta2}
&= \beta V_{j {\SB n}}{\SB \lambda_{n}^{-1}} [V^{-1}]_{n {\SB m}} \langle \delta N_{{\SB m}}\delta N_{\ell}\rangle \ .
\end{align}
For the case of a zero eigenvalue{\SB, $\lambda_n = 0$, and $\lambda_n^{-1}$ is undefined, seemingly indicating that the integral in \eqref{zeta} does not converge; however, an ergodic stationary process has an autocovariance that does not contain any time-independent elements~\cite{Gardiner}, thus all $\lambda_n = 0$ components cancel in the product $V_{j n}e^{-\lambda_{n}t}[V^{-1}]_{nm}\langle \delta N_{m}\delta N_{\ell}\rangle$, and the integral converges.}

A \emph{conversion network} allows only conversion, degradation, and source reactions~\cite{Firstorderreactions}. It is \emph{open} when it has at least one degradation or source reaction. 
The equilibrium distribution (reached in the $t\to \infty$ limit of \eqref{Mean time evolution}) of any species in an open linear conversion network is a Poisson distribution, with mean and covariance~\cite{Firstorderreactions} 
\begin{align}
\label{Open Variance}
{\SB \langle \delta N_{m}
^2\rangle^{\rm o} = \langle N_{m} \rangle^{\rm o} = 
 V_{mn}\lambda_{n}^{-1} [V^{-1}]_{nj}k^{\rm s}_{ j} } \ ,
\end{align}
{\SB and $\langle \delta N_{m}\delta N_{\ell} \rangle^{\rm o} = 0$ if $m\neq\ell$.}

The friction tensor for an open system can therefore be fully determined from the equilibrium mean and reaction rates as
\begin{align}
\label{Zeta open}
\zeta_{j\ell}^{\rm o} = 
\beta V_{j {\SB  n}}{\SB \lambda_n^{-1}}[V^{-1}]_{n {\SB m}}\langle N _{ {\SB m}}\rangle^{\rm o} \delta_{\SB m\ell} \ ,
\end{align}
{\SB where $\delta_{m\ell}$ is the Kronecker delta, equal to $1$ if $m = \ell$, and $0$ otherwise.}
The relaxation time is $\tau_{j\ell}^{\rm o} = 
V_{j{\SB n} } {\SB \lambda_n^{-1}} [V^{-1}]_{n\ell}$, which is proportional to the mean copy number~\eqref{Open Variance}. Hence an increase in mean copy number has the compound effect of increasing both the size and lifetime of fluctuations. Therefore, the designed chemical-potential protocol drives slowly in areas of large mean copy number and quickly in areas of low mean copy number.

For a linear \emph{closed} conversion network (no sources or degradation), the equilibrium distribution is not Poisson~\cite{Firstorderreactions}, but the mean, variance, and covariance can still be solved analytically using standard linear algebra techniques~\cite{Closedfirstorder,Firstorderreactions}. The equilibrium covariance is
\begin{align}
\label{closed covariance}
\langle \delta N_{{\SB m}}
\delta N_{\ell}\rangle^{\rm c} = 
\begin{cases}
{\SB \langle N_{m}\rangle^{\rm c}\left(1-\frac{\langle N_{\ell}\rangle^{\rm c}}{\Ntot}\right) } \ &{\SB , \quad \ell = m } \\
{\SB -\frac{\langle N_{m}\rangle^{\rm c}\langle N_{\ell}\rangle^{\rm c}}{\Ntot} \ } & {\SB , \quad \ell \neq m }
\end{cases}
\end{align}
where $\Ntot = \sum_{j}N_{j}$ is the total number of chemical molecules. For chemical reaction systems with a \emph{strongly connected} reaction graph (\emph{i.e.}, any species can be reached from any other via a set of allowed reactions), $K$ has exactly one zero eigenvalue, and the equilibrium probability distribution across all species is multinomial~\cite{Firstorderreactions}, $\pi_{j} = v^{0}_{j}/\sum_{\ell}v^{0}_{\ell}$, where $v^{0}_{j}$ is the $j$th component of the eigenvector with zero eigenvalue.
The multinomial mean copy number of species $j$ is simply $\langle N_{j} \rangle = \Ntot\pi_{j}$, producing covariance
\begin{align}
\label{closed correlation}
\langle \delta N_{{\SB m}}
\delta N_{\ell}\rangle^{\rm c} = 
\begin{cases}
{\SB \Ntot\pi_{m}\left(1-\pi_{\ell}\right) \ }&{\SB , \quad \ell = m }\\
{\SB -\Ntot \pi_{m} \pi_{\ell} \ }&{\SB , \quad \ell \neq m}
\end{cases}
\end{align}

Substituting the covariance~\eqref{closed covariance} into the friction~{\SB\eqref{zeta2}} gives
\begin{align}
\label{Closed Zeta2}
\zeta_{j \ell}^{\rm c} = 
&\beta V_{j {\SB n}}{\SB \lambda_{n}^{-1}} [V^{-1}]_{n {\SB m}} \langle N_{{\SB m}}\rangle^{\rm c}\left(\delta_{{\SB m} \ell}-\frac{\langle N_{\ell}\rangle^{\rm c}}{\Ntot}\right) \ .
\end{align}
Unlike for the open system, the closed covariance~\eqref{closed covariance} does not monotonically increase with mean copy number, but rather is largest when the two species have equal mean copy numbers and is smallest when one species dominates. If $m = \ell$, then the covariance reduces to the variance, which is maximized at $\langle N_{m}\rangle^{\rm c} = \Ntot/2$ and minimized at $\langle N_{m}\rangle^{\rm c} = \Ntot$ or $\langle N_{m} \rangle^{\rm c} = 0$. When $m\neq \ell$, $\langle \delta N_{m}\delta N_{\ell}\rangle^{\rm c} = -\langle N_{m}\rangle^{\rm c}\langle N_{\ell}\rangle^{\rm c}/\Ntot$, which is always negative and reaches its maximum magnitude when $\langle N_{m}\rangle^{\rm c} = \langle N_{\ell} \rangle^{\rm c} = \Ntot/2$. 

For small mean copy number relative to the total, $\langle N_{\ell}\rangle^{\rm c} \ll \Ntot$, the friction of a closed system~\eqref{Closed Zeta2} reduces to that of an open system~\eqref{Zeta open}, since the second term in parentheses in \eqref{Closed Zeta2} becomes negligible. The large total number of molecules acts as a constant source, or chemical bath, making the closed and open systems equivalent.

In order to interpret the form of the closed-system relaxation time $\tau_{j\ell}^{\rm c} = V_{j{\SB n}} {\SB \lambda_n^{-1}} [V^{-1}]_{n\ell}$, we recognize that the eigenvalues of $K$ in a closed system have non-negative real components~\cite{Firstorderreactions}. Furthermore, if the system satisfies detailed balance, then the eigenvalues of $K$ are real~\cite{realeigenvalues,realeigenvaluestwo}. Thus $\tau_{j \ell}$ is non-negative. As we have seen, all off-diagonal components of the covariance are negative and all diagonal components are positive, therefore the same is true of the friction tensor, the product of covariance and relaxation time. Although the friction tensor can have negative specific entries, it is positive semidefinite since it is an auto-covariance matrix~\cite{OptimalPaths}.
	
The friction tensors \eqref{zeta}, \eqref{Zeta open}, and \eqref{Closed Zeta2} imply analytic solutions for the designed protocol of any linear Markov chemical reaction. In the following sections we examine specific reaction networks to gain further insight into designed protocols.

\section{\label{Closed system}Closed system}
\begin{figure}
\begin{center}
\includegraphics[scale = 0.25]{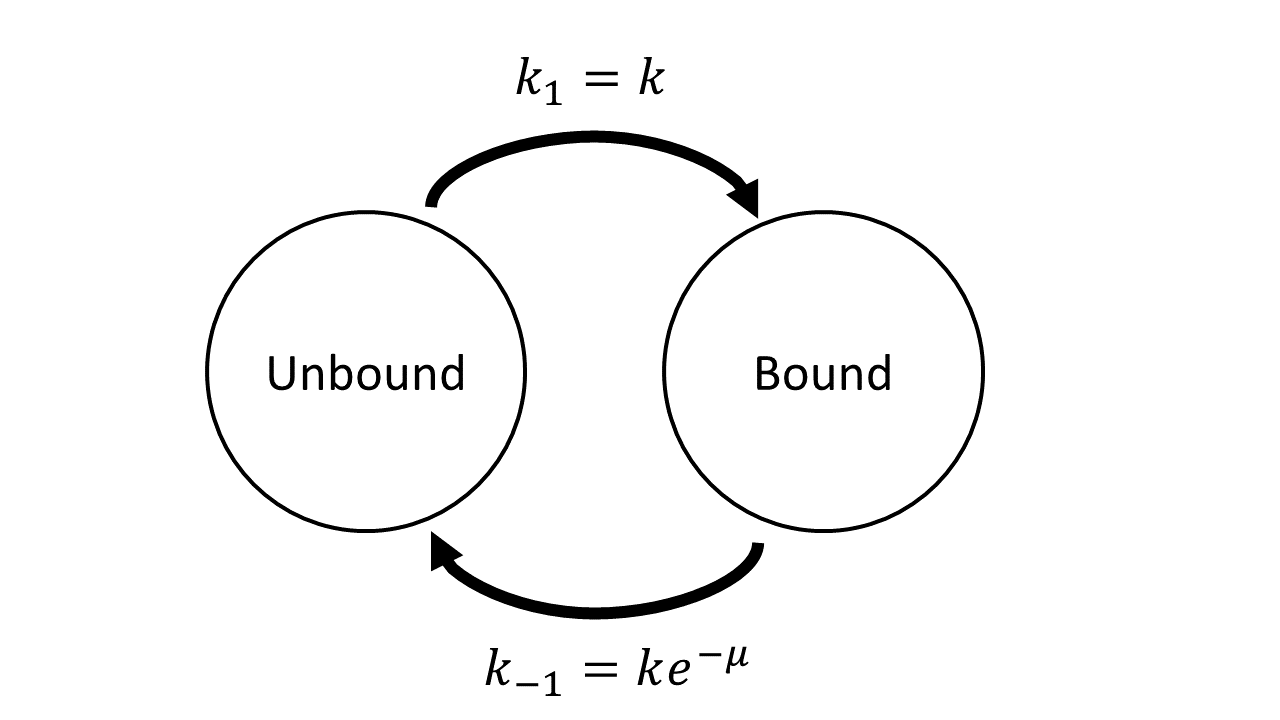}
\caption{Two-state chemical reaction network representing bound and unbound G-proteins. Proteins bind at rate $k_1=k$ and unbind at rate $k_{-1}=ke^{-\mu}$.}
\label{Reaction}
\end{center}
\end{figure}

As a simple tractable model, we examine a two-state chemical reaction with respective binding and unbinding rates $k_1$ and $k_{-1}$ (Fig.~\ref{Reaction}), nominally meant to represent G-proteins binding to the GPCR at the cell membrane. 
	
{\SB In this model, the signal is the chemical potential difference between the unbound and bound states, which regulates the number of bound G-proteins. In the unbound state, the G-proteins are active, leading to downstream reactions resulting in the cellular response.} The chemical potential is the externally controlled signal, for example as modulated by the number of expressed agonist molecules{\SB. In this model, the chemical potential regulates the number of unbound (active) G-proteins and hence the cellular response.}
	
It is natural to model the membrane binding rate $k_1=k$ as depending on the dynamic encounter rate and not on the strength of the chemical potential, and the membrane unbinding rate $k_{-1}$ as depending on how tightly the protein is bound, and hence on the chemical potential difference $\mu$ between unbound and bound states, as
\begin{equation}
\label{rate chemical dependence}
k_{-1} = ke^{-\mu} \ .
\end{equation}
$\mu = 0$ produces equal binding and unbinding rates, $k_1 = k_{-1}$. (This specific dependence of rates on chemical potential is consistent with~\cite{brownsivak2017,brownsivak2018} for a splitting factor~\cite{split1,split2} of $0$, although our framework could be applied to any splitting factor.) For simplicity, here and in subsequent sections, energies are written in units of $k_{\rm B}T$ (equivalent to setting $\beta = 1$).
	
We additionally assume a fixed total number of molecules $\Ntot = \Nub+\Nb$, with variable numbers of unbound ($\Nub$) and bound ($\Nb$) molecules. The reaction-rate matrix is
\begin{align}
K =	\left[ {\begin{array}{cc}
	ke^{- \mu}&-k \\
	-ke^{- \mu}&k \\
	\end{array} } \right] \ .
\end{align}
	
\begin{figure}
\includegraphics[width=\columnwidth]{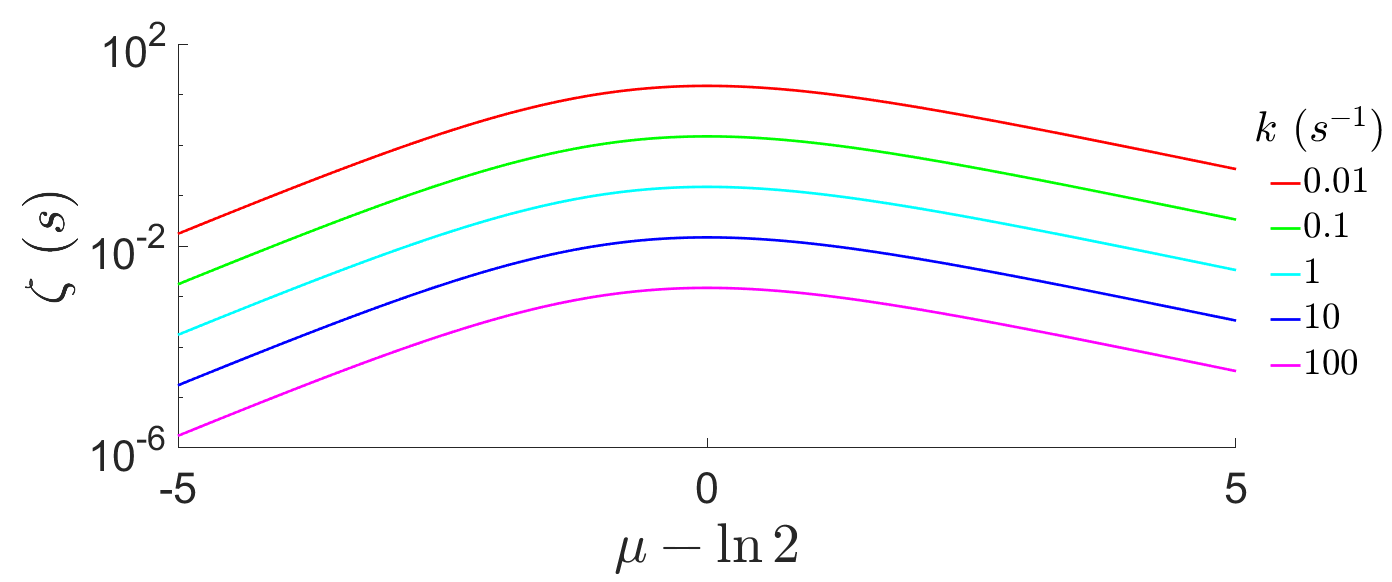}
\caption{Generalized friction coefficient $\zeta$ (in units of seconds, since $k_{\rm B}T$ is set to unity) as a function of chemical potential $\mu$, for various binding rates $k$ (different colors). The horizontal axis is shifted by $\ln 2$ so that the friction of the closed system is maximized at $0$. For simplicity, the total protein copy number $\Ntot$ is normalized to $1$.}
\label{friction Plot}
\end{figure}

In \S\ref{Linear Markov chemical reaction networks}, we derived simple expressions for the auto-covariance~\eqref{autocovariance}, equilibrium covariance~\eqref{closed correlation}, and friction~\eqref{Closed Zeta2}. With one chemical potential, there is only the $j=\ell=1$ component, giving equilibrium variance
\begin{align}
\label{closedvar}
\langle (\delta \Nb)^{2}\rangle^{\rm c}_{\mu} = \Ntot \frac{e^{-\mu}}{(1+e^{-\mu})^2} \ ,
\end{align}
relaxation time
\begin{align}
\label{closedtau}
\tau(\mu) = \frac{1}{k(1+e^{-\mu})} \ ,
\end{align}
and friction
\begin{align}
\label{closed friction}
\zeta(\mu) &= \Ntot \frac{e^{-\mu}}{k(1+e^{-\mu})^3} \ .
\end{align} 
The variance is maximized at $\mu = 0$. 
For $e^{\mu} \gg 1$, the variance decays exponentially with $\mu$ as $\langle (\delta \Nb)^{2}\rangle^{\rm c}_{\mu} \approx \Ntot e^{-\mu}$.
Figure~\ref{friction Plot} plots the dependence of friction coefficient on $\mu$, for several binding rates $k$. 
 
Physically, as $\mu$ increases, molecules are held more tightly to the membrane (unbinding rate decreases), and thus copy-number fluctuations relax more slowly. The relaxation time is sigmoidal in $\mu$, with $\tau(\mu \to -\infty) \to 0$ and $\tau(\mu \to \infty) \to 1/k$. The first limit corresponds to molecules bound very loosely to the membrane, such that the unbinding rate is much larger than the binding rate, with fluctuations decaying rapidly. The latter limit corresponds to tightly bound molecules such that the binding rate is much larger than unbinding, causing fluctuations to decay slowly and most molecules to be bound: the relaxation time is maximized when all molecules are bound. Ultimately, this asymmetry in relaxation time is caused by the asymmetric dependence of the forward and reverse reaction rates on chemical potential: $k_1$ is independent of $\mu$ and $k_{-1}\propto e^{-\mu}$.
 
The friction is minimized (and vanishes) when either all molecules are bound or all are unbound. The friction peaks at $\mu = \ln2$, when $2/3$ of all molecules are bound, $\langle \Nb \rangle^{\rm c} = \tfrac{2}{3}\Ntot$. Physically, the resistance increases when driving away from either all-bound or all-unbound: as the mean copy number of the less common species increases, the resistance to changes in chemical potential increases. This can be rationalized because the variance is maximized at $\mu=0$, when each state (bound and unbound) contains on average half the total number of molecules, whereas the relaxation time is maximized when all the molecules are bound, thus shifting the maximal friction to occur past an even split in each state. At chemical potentials well below this maximum (for $e^{\mu} \ll 1$), the friction increases as $e^{2\mu}$, whereas for large chemical potentials ($e^{\mu} \gg 1$), the friction decays exponentially with chemical potential, $\zeta \to e^{-\mu}$. Figure~\ref{friction Plot} shows these differing slopes. 
 
The designed protocol drives slowly in control parameter regimes of high friction which, due to the exponential dependence of friction on chemical potential~\eqref{closed friction}, produces large variations in chemical potential velocity and potentially large energetic saving. This behaviour is illustrated in Fig.~\ref{friction Plot}.
 
With a single control parameter, the designed protocol is easily solved using \eqref{lambdaoptdot2}:
\begin{align}
\label{closed lambdaoptdot3}
\md_t \mu^{\rm des}|_{\mu} 
= &\frac{ 2\sqrt{1+e^{\mu}}(1+e^{-\mu})}{\Delta t} \left(\frac{1}{\sqrt{1+e^{\muI}}}-\frac{1}{\sqrt{1+e^{\muF}}}\right) \ .
\end{align}
The velocity of the designed protocol reaches a minimum when the friction is at a maximum, $\mu = \ln 2$. Appendix~\ref{Designed copy number} derives the equivalent designed mean-copy-number protocol, which increases as $\md_t \langle \Nb\rangle^{\rm c~des} \propto \sqrt{\langle \Nub\rangle^{\rm c}}$. Appendix~\ref{initial and final velocity} compares the initial and final designed protocol velocities, and demonstrates that for small changes in chemical potential, the designed protocol reduces to the naive.
	
The designed protocol produces an excess work 
\begin{align}
\label{closeddesignedwork}
W_{\rm ex}^{\rm c~des} = &\frac{4\Ntot }{k \Delta t} \left( \frac{1}{\sqrt{1+e^{\muI}}} - \frac{1}{\sqrt{1+e^{\muF}}}\right)^{2} \ .
\end{align}
For significant changes in chemical potential, either increases ($e^{\muF} \gg e^{\muI}$ and $e^{\muF} \gg 1$) or decreases ($e^{\muF} \ll e^{\muI}$ and $e^{\muF} \ll 1)$, the designed excess work becomes independent of $\muF$.
	
The naive protocol changes chemical potential at constant velocity $\md_t \mu^{\rm naive} = \Delta\mu/\Delta t$ and produces excess work~\eqref{excess work}
\begin{align}
\label{closednaivework}
W_{\rm ex}^{\rm c~naive} = \Ntot \frac{\Delta\mu}{\Delta t} 
\frac{1}{2k}\left[\frac{1+2e^{\muI}}{(1+e^{\muI})^{2}}-\frac{1+2e^{\muF}}{(1+e^{\muF})^{2}}\right] \ .
\end{align}
{\SB A linear protocol represents the conceptually simplest one for comparison and a natural choice in the absence of any other information about how to proceed.}  
For significant changes in chemical potential, the naive excess work \eqref{closednaivework} scales linearly with $\Delta\mu \equiv \muF-\muI$. 
This is in contrast to the excess work from the designed protocol \eqref{closeddesignedwork}, which becomes independent of $\muF$ in this limit.

We quantify the thermodynamic benefit of designed driving by the ratio of the excess works incurred during the naive and designed protocols~\eqref{Ratio}:
\begin{align}
\label{closedratio}
\frac{W_{\rm ex}^{\rm c~naive}}{W_{\rm ex}^{\rm c~des}} &= \Delta\mu
\frac{\frac{(1+2e^{\muI})(1+e^{\muF})}{1+e^{\muI}}-\frac{(1+2e^{\muF})(1+e^{\muI})}{1+e^{\muF}}}{8\left(\sqrt{1+e^{\muI}}-\sqrt{1+e^{\muF}}\right)^{2}} \ . \end{align}
The ratio does not depend on the bare binding/unbinding rate $k$. For significant chemical potential changes, the excess-work ratio scales linearly with $\Delta\mu$. Appendix~\ref{small chemical potential} shows that for small changes $\Delta\mu$ in chemical potential, both the naive excess work and the excess-work ratio increase quadratically in $\Delta\mu$.

\begin{figure}
\includegraphics[width=\columnwidth]{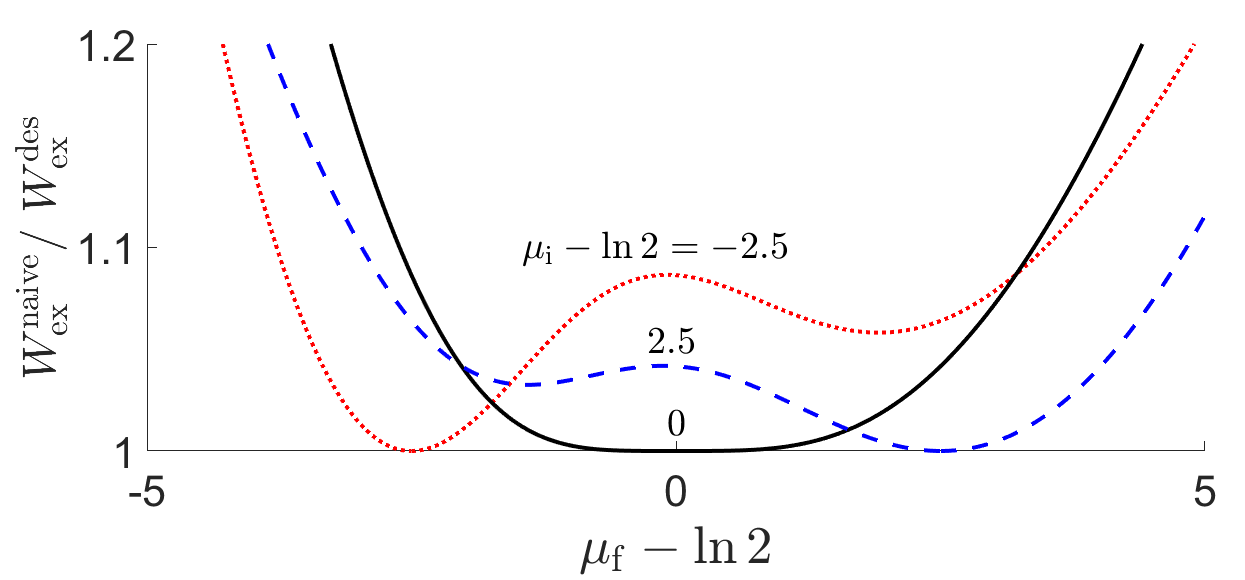}
\caption{The ratio of naive to designed excess works as a function of the final chemical potential $\muF - \ln 2$, for varying shifted initial chemical potential $\muI - \ln 2$ (different colors). Horizontal axis is shifted to $\muF - \ln 2$, so that the protocol crosses the maximal friction at $0$.}
\label{Ratio Plot}
\end{figure}

The only parameters in \eqref{closedratio} are the initial and final chemical potentials $\muI$ and $\muF$. Figure~\ref{Ratio Plot} demonstrates that the excess work ratio is non-monotonic in $\muF$, empirically peaking near the local maximum in the friction; however, after decreasing for a short distance, the ratio begins to increase linearly. This transition can occur for either positive or negative chemical potential distances, depending on which side of the maximum friction the protocol starts. Such a feature is not found for a protocol initially at the peak friction. The asymmetry in excess work ratio on different sides of the maximal friction is caused by the friction scaling as $e^{2\mu}$ for chemical potentials below the peak and as $e^{-\mu}$ for chemical potentials above the peak (Fig.~\ref{friction Plot}), itself a result of the asymmetric chemical potential dependence of the forward and reverse reaction rates. Outside of this region, more significant chemical potential changes still produce greater benefits from the designed protocol (quadratic for small $\Delta\mu$ and linear for large $\Delta\mu$).
 
\section{\label{two state open}Open system}
When the unbinding rate is much larger than the binding rate (for $e^{\mu} \ll 1$), and hence $\langle \Nub\rangle^{\rm c} \gg \langle \Nb\rangle^{\rm c}$, $\Nub$ is effectively constant over copy-number fluctuations, and thus the system is effectively open, with $K = -K^{\rm d} = -k_{-1}$ and $k^{\rm s} = \Nbath k$. This limit produces particularly simple forms for
the variance~\eqref{closedvar}
\begin{align}
\label{openvar}
\langle \left(\delta N_{B}\right)^{2}\rangle_{\mu}^{\rm o} = \Nbath e^{\mu} \ ,
\end{align}
relaxation time~\eqref{closedtau}
\begin{align}
\label{opentau}
\tau(\mu) = \frac{e^{\mu}}{k} \ ,
\end{align}
and friction~\eqref{closed friction}
\begin{align}
\label{openfriction}
\zeta(\mu) = \Nbath\frac{e^{2 \mu}}{k} \ .
\end{align}

Both the copy-number variance~\eqref{openvar} and relaxation time~\eqref{opentau} increase exponentially with $\mu$. 
The relaxation time only depends on the unbinding rate, the characteristic time for a membrane-bound molecule to unbind, and since the (Poissonian) copy-number variance equals the mean, larger $\mu$ decreases the unbinding rate, increasing copy-number mean and thus decreasing the relaxation time and variance.

Combining \eqref{lambdaoptdot2} with \eqref{openfriction} leads to the designed protocol velocity,
\begin{equation}
\label{muoptdot open}
\md_t \mu^{\rm des}|_{\mu} = 
\frac{e^{-\mu}\left(e^{\muF}-e^{\muI}\right)}{\Delta t} \ .
\end{equation}
When driving the system from low to high chemical potential, as time progresses the designed protocol slows as $e^{-\mu}$. Appendix~\ref{Designed copy number} derives the designed protocol in terms of mean copy number, which amounts to driving at constant velocity $\md_t \langle \Nb\rangle^{\rm o} = \Delta \langle \Nb\rangle^{\rm o}/\Delta t$, equivalent to the naive mean-copy-number protocol. Appendix~\ref{initial and final velocity} shows that the initial velocity is exponentially faster than the final, and for small changes in chemical potential the designed protocol reduces to the naive. 

The designed chemical-potential protocol produces a constant excess power, leading to total excess work~\eqref{excess work}
\begin{align}
\label{opendesignedwork}
W_{\rm ex}^{\rm o~des} = \Nbath 
\frac{e^{2 \muI}}{k \Delta t}(e^{\Delta\mu} - 1)^2 \ .
\end{align}
For large increases in chemical potential ($e^{\Delta\mu} \gg 1$), the designed excess work increases exponentially in chemical potential distance, incurring large energetic costs; conversely, for large decreases in chemical potential, the excess work is independent of the chemical potential change $\Delta \mu$.

The excess power during the naive (constant-velocity) protocol \eqref{LR excess power} produces excess work~\eqref{excess work}
\begin{align}
\label{opennaivework}
W_{\rm ex}^{\rm o~naive} = \Nbath \frac{\Delta\mu}{\Delta t} \frac{ e^{2 \muI}}{2k}(e^{2\Delta\mu} - 1) \ .
\end{align}
For large $\Delta\mu$, the naive excess work increases exponentially in chemical potential, thus incurring huge energetic costs. When significantly reducing chemical potential ($e^{2\Delta\mu} \ll 1$), the excess work increases linearly with decreasing $\Delta\mu$, which is a significantly slower rate than for chemical potential increases, but still significantly faster than the designed protocol~\eqref{opendesignedwork}, for which the excess work becomes independent of chemical potential. The friction is smaller at lower chemical potentials; therefore, reducing chemical potential carries the system through regions of control parameter space with lower resistance, thereby slowing the increase in energetic cost associated with greater-magnitude changes of chemical potential. Increasing chemical potential carries the system towards parameter space with higher resistance, further exacerbating the energetic cost. 

The excess work ratio is
\begin{align}
\frac{W_{\rm ex}^{\rm o~naive}}{W_{\rm ex}^{\rm o~des}}= 
\frac{\Delta\mu}{2}\frac{e^{\Delta\mu} + 1}{e^{\Delta\mu}-1} \ .
\end{align}
Despite the magnitude of the naive work increasing slowly for chemical potential reductions, the ratio is symmetric about $\Delta\mu = 0$. As the chemical potential change $|\Delta\mu|$ increases, so does the ratio of the excess works, and hence the energetic savings from using the designed protocol.

\section{Discussion}
Living things accrue a selective advantage if they can use less energy to achieve their required functions. In the task of dynamic cell-cell signaling, methods for achieving given changes in the target cell at minimum energy expenditure may point toward design principles for intercellular communication. 
 
We have {\SB adapted a theoretical framework for a novel problem domain,}
to approximate the energetic cost of rapidly changing chemical potential, and we used it to design finite-time chemical-potential protocols that (under linear response) reduce the excess work incurred in dynamically driven biochemical reaction networks. We analyzed the designed protocol for an arbitrary linear Markov chemical reaction network, and we applied it to an exactly solvable model system with only binding/unbinding reactions: a closed system with a fixed total number of proteins, which in the limit of small chemical potential can effectively be treated as an open system connected to a chemical bath. The designed protocol for such a linear chemical reaction system is simply determined by the collection of reaction rates. This approach can be generalized to non-linear chemical reactions by using moment-closure techniques to obtain approximate solutions.
	
We find that for a two-state closed system, the generalized friction---the resistance to changes in chemical potential---is minimized (at $0$) when all proteins are either bound or unbound, and is maximized when $2/3$ of all proteins are bound, when the binding rate equals twice the unbinding rate. This corresponds to a balance between the largest fluctuations (when the binding rate equals the unbinding rate) and the largest relaxation time (for small unbinding rate and tightly bound proteins). Under these conditions, the designed protocol changes the chemical potential slowest at intermediate mean copy number. For an open system, the friction increases monotonically with mean copy number. Therefore, a protocol that minimizes energetic cost (near equilibrium) changes the chemical potential slowly when mean copy number is high and quickly when mean copy number is low. 
	
Similar analysis shows that when chemical potential exponentially enhances binding rather than exponentially suppressing unbinding (for a splitting factor~\cite{split1,split2} of $1$), friction is maximized when $1/3$ of all proteins are bound, corresponding to a binding rate half of the unbinding rate. When the chemical potential enhances binding and suppresses unbinding equally (splitting factor of $1/2$) friction is maximized when $1/2$ of all proteins are bound, corresponding to equal binding and unbinding rates; however, no closed-form solutions for the designed protocols and excess works for intermediate splitting factors in $(0,1)$ are known.

Our analysis focused on chemical networks with known (and simple) topologies and reaction rates. It would be interesting to see how these results change for more complicated chemical networks. For example, a chemically bistable system (with two metastable copy-number states) would have significantly longer relaxation times at chemical potentials for which the system is bistable. Similar to recent results for a particle diffusing over a bistable potential~\cite{Workratio}, we expect the friction to be peaked at such bistability-inducing chemical potentials, meaning that work-minimizing protocols slow down near the threshold chemical potential to allow chemical fluctuations time to kick the system into the desired metastable state. 

In the absence of such detailed information, one could phenomenologically map out the generalized friction coefficient through monitoring copy-number fluctuations~\cite{Dar:2012ab} at various fixed chemical potentials, then use the linear-response theory to infer the corresponding designed protocols, in analogy to recent work in single-molecule contexts~\cite{tafoya2018using}. 

{\SB Although our study is presented in the context of cell-cell signaling, our results hold for more general chemical reaction systems. Traditional
stochastic thermodynamics treatments of chemical reaction networks~\cite{Gaspard2004,Andrieux2004,Andrieux2007,Vellela2008,Ge2009} feature sustained chemical currents at fixed chemical potentials. In contrast, our setup dynamically varies chemical potential~\cite{Schmiedl2007,Rao2016,Falasco2018,Rao2018}, with our \eqref{Free Energy Work} corresponding to a linear response approximation to the ``driving work''~\cite{Rao2018}. One major benefit of this approximation is that it gives a relatively straightforward prescription for designing protocols that reduce dissipative work.

In general, thermodynamic consistency demands an accounting of the dissipative costs associated with implementing a particular time-asymmetric, detailed-balance breaking protocol~\cite{machta15,BryantMachta}. However, that contribution scales sub-extensively with system size, whereas the frictional dissipation modeled here scales extensively, so should dominate for larger systems such as a collection of cells each with many receptors. In the interests of a simple and tractable model system, we here focused on the frictional dissipation.}
 
The less energy used during operation, the fewer signaling proteins that must be produced and dynamically secreted. Such designed control analysis makes strong predictions about the dynamic interactions that communicate information and regulate behavior in an energetically efficient manner. To the extent that energetic efficiency is an important functional characteristic for such signaling pathways, experiments may uncover signatures of these design criteria in evolved molecular and cellular systems.
 
There are several known mechanisms by which a signaling cell can dynamically control a target cell's response to take advantage of designed protocols. The simplest method is by dynamically controlling the number of agonists secreted. Another method, used by $\beta$-adrenergic receptor kinases~\cite{nobel80} and rhodopsin kinase~\cite{nobel81}, is phosphorylation, which increases the affinity of the receptor for regulatory proteins called arrestins~\cite{nobel82,nobel83}, in turn down-regulating the number of active receptors. Additionally, recycling of receptors and internalization via endocytosis can regulate the signal~\cite{nobel84,nobel85}. All of these techniques are employed to adjust the number of active GPCRs and therefore allow for the control of the binding affinity and reaction rates of the G-protein between the bound and unbound states.
	
Recent experimental advances make possible the precise spatial and temporal control of binding affinity between different chemical species, and hence of protein spatial localization within a cell. In particular, optogenetic techniques allow for the use of light to adjust the binding affinity between a light-gated protein and its binding partner~\cite{accuracy}. Changes in binding affinity are effectively changes in the chemical potential of one class of proteins in the vicinity of another, thus allowing for the dynamic experimental implementation of our proposed control strategies. 
{\SB Quantitative fluorescence microscopy techniques~\cite{QuantFluorMic} could permit quantification of the actual nonequilibrium changes in protein copy numbers, and thus of the dissipative chemical-potential work and the ability of such protocols to achieve desired downstream changes.}

\begin{acknowledgments}
The authors thank Aidan I.\ Brown, Joseph N.\ E.\ Lucero, and Alzbeta Medvedova (SFU Physics) for enlightening discussions. This research was supported by funding from a Natural Sciences and Engineering Research Council of Canada (NSERC) Discovery Grant, a Tier-II Canada Research Chair, and the Faculty of Science, Simon Fraser University through the President's Research Start-up Grant (all to D.\ A.\ S.).
\end{acknowledgments}

\appendix
{\SB
\section{\label{Exact Work}Exact work}
In the GCE, the composition of the system can change by adding or removing particles through, for example, a chemical reaction. To account for this, we identify the total power as the sum of the mechanical and chemical powers
\begin{align}
\md_t W = \md_t W_{\rm mech} + \md_t W_{\rm chem} \ ,
\end{align}
where the mechanical power is the
nonequilibrium average of the change in internal energy along the control parameter protocol $\Lambda$,
\begin{align}
&\md_t W_{\rm mech} = \langle \partial_{\lambda_{j}}U \rangle_{\Lambda} \ \md_t \lambda_{j} \ ,
\label{eq:dWmech}
\end{align}
and the chemical power 
is due to
the change in composition,
\begin{align}
&\md_t W_{\rm chem} = -\langle \partial_{\lambda_{j}}(\mu_{\ell}N_{\ell}) \rangle_{\Lambda} \ \md_t \lambda_{j} \ .
\end{align}
Combining the mechanical and chemical contributions gives 
the natural extension of \eqref{eq:dWmech} to the GCE,
the total power
\begin{align}
\md_t W = \langle \partial_{\lambda_{j}}U \rangle_{\Lambda} \ \md_t \lambda_{j} -\langle \partial_{\lambda_{j}}(\mu_{\ell} N_{\ell}) \rangle_{\Lambda} \ \md_t \lambda_{j} \ ,
\label{eq:dWGCE}
\end{align}
with corresponding conjugate force $\langle f \rangle_{\lambda} \equiv -\partial_{\lambda}\Phi_{\rm G}$.
When the control parameter is $\lambda_j=\mu_j$, the energy $U$ and copy number $N_j$ are independent of $\lambda$, so \eqref{eq:dWGCE} reduces to the average instantaneous change in excess chemical work along a particular chemical-potential protocol $M$,
\begin{align}
\md_t W_{\rm ex} = -\langle \delta N_{j} \rangle_{M} \, \md_t \mu_{j} \ ,
\label{CP exact power}
\end{align}

This definition is consistent with the recently defined driving work in Rao, Falasco, \& Esposito~\cite{Rao2016,Falasco2018,Rao2018}, where our largest indivisible units are the chemical species $N_{j}$ (as opposed to their chemical moieties).
In our case we assume no sustained chemical currents (zero non-conservative work), so the only dissipative contribution is the driving work.

\section{\label{Linear Response Approximation}Linear-response approximation}
There are two approximations leading to the friction-tensor formulation of the excess chemical work in \eqref{LR excess power}. 
One is the linear-response approximation: over time scales where the response function $\md_t\langle \delta f(t)\delta f(0)\rangle_{\lambda(t_0)}|_{\tau}$ is significantly different from zero, both the nonequilibrium response $\langle \Delta f(t_0)\rangle_{\Lambda} \equiv f(t_0) -\langle f \rangle_{\lambda(t_0)}$ (the deviation of the conjugate force $f$ at time $t_0$ from the average conjugate force $f$ at equilibrium under control parameter $\lambda(t_0)$) and the equilibrium change $\langle f(t_0)\rangle - \langle f(t_0-\tau)\rangle$ can be Taylor expanded to first order in the control parameter change $\lambda(t_0)-\lambda(t_0-\tau)$.
The other is smooth protocols: the control parameter can be Taylor expanded to zeroth order, $\dot{\lambda}(t_0-t'') = \dot{\lambda}(t_0)$, when the control protocol $\Lambda$ is sufficiently smooth, such that $\dot{\lambda}^j(t) \gg (t'-t) \ddot{\lambda}^j(t)$ for time separations $t'-t$ over which the conjugate force autocorrelation $\langle \delta f_i(0)\delta f_j(t-t')\rangle_{\lambda(t)}$ is significantly greater than zero, i.e., over time scales less than the relaxation time of the conjugate forces.

As a direct test of the linear-response approximation, we calculate the exact excess power~\eqref{CP exact power} exactly for the two-state closed system with control parameter $\mu(t)$, giving
\begin{align}
\md_t\langle \Nb \rangle_{M} = k\left[\Ntot -\langle \Nb \rangle_{M} (e^{-\mu(t)}+1)\right] \ ,
\label{changeinN}
\end{align}
subject to an equilibrium initial condition.
The exact excess power~\eqref{LR excess power} is obtained by solving this numerically for a given control parameter protocol and subtracting the equilibrium mean 
\begin{align}
\langle \Nb \rangle_{\mu} = \frac{\Ntot}{1+e^{-\mu}}
\end{align}
at each chemical potential $\mu$ along the protocol.

Figure~\ref{Power Plot} compares numerical solutions for the exact excess power~\eqref{CP exact power} with analytic approximate solutions for the naive (constant-velocity) and designed~\eqref{closed lambdaoptdot3} protocols. As the average driving velocity $c\equiv\Delta\mu /\Delta t$ decreases, the exact solutions converge to the approximate result, which has a maximum at $\mu = \ln 2$ for the naive protocol and is constant for the designed protocol. It is not until the driving speed is roughly the same speed as the bare reaction rate, $|c|\approx k$, that the exact result significantly deviates from the approximation. In general, for chemical reactions that take place on short time scales (large $k$) the approximation should be valid; however, the exact speed at which it deviates significantly will depend on the specifics of the reaction network. 

While the approximate excess power is always independent of the initial chemical potential and protocol direction (forward or reverse), for the exact calculations this is noticeably violated for fast driving ($|c|/k \gtrsim 1$). For the naive protocol, the excess power peaks after passing the maximal friction at $\mu =\ln 2$. Intuitively, as the driving speed increases, the system state (mean copy number of bound proteins) increasingly lags behind the equilibrium value. If we assume the mean-variance relation~\eqref{closed covariance} still holds even though the nonequilibrium mean lags the corresponding equilibrium mean (amounting to an assumption of \emph{endoreversibility}~\cite{Salamon:1983:PhysRevLett}), this shifts the maximum variance (and hence maximum excess power in naive protocols) to larger (smaller) chemical potentials for the forward (reverse) protocols. Since the designed protocol slows down where the equilibrium friction coefficient is largest, the lag---and concomitant shifting of the variance maximum to later in the protocol---means that the designed protocol slows down too early, and then speeds up too early. This produces the rapid increase in excess power ($c/k = 10$ curves in Fig.~\ref{Power Plot}) late in the protocol.

\begin{figure}
\includegraphics[width=\columnwidth]{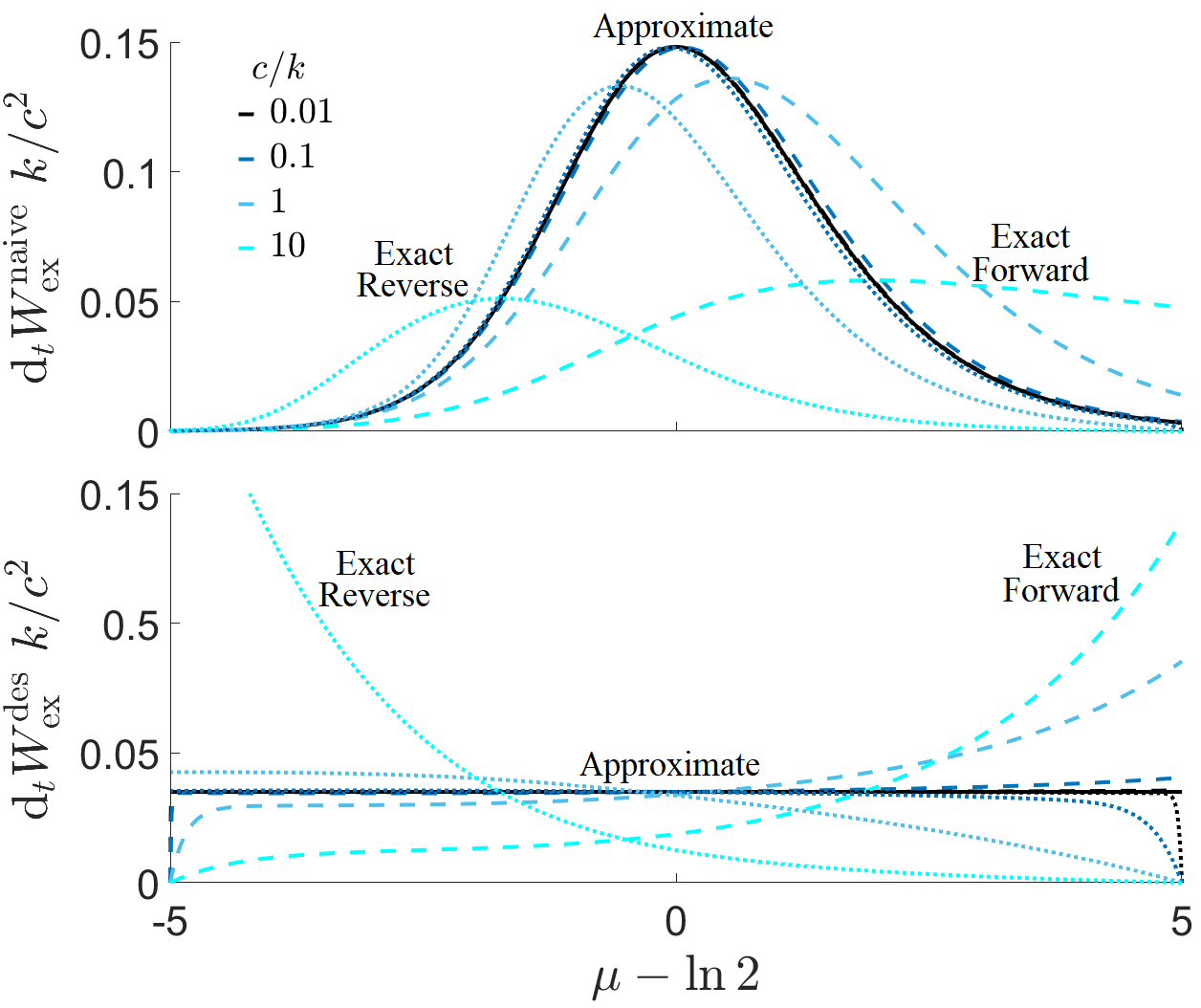}
\caption{{\SB Excess power ${\rm d}_{t} W_{\rm ex}$ along the naive (top) and designed (bottom) protocols as a function of chemical potential $\mu$. Analytic approximation~\eqref{LR excess power} (solid black curve) and exact calculation~\eqref{CP exact power} for various average chemical potential velocities $c \equiv \Delta\mu/\Delta t$ (different colors) for forward (dashed) and reverse (dotted) protocols, starting at $\mu-\ln 2 = -5$ and $5$ respectively. The horizontal axis is shifted by $\ln 2$ so the friction is maximized at $0$. The driving velocity is scaled by the relevant reaction time scale $k$ (excess power only depends on the ratio $c/k$) and the excess power is scaled such that in the limit of slow driving the exact solutions collapse onto a single curve, the analytic approximation. For simplicity, the total protein copy number $N_{\rm tot}$ is normalized to $1$.}}
\label{Power Plot}
\end{figure}
}

\section{\label{Designed copy number}Designed mean-copy-number protocol}
In the grand canonical {\SB ensemble} (GCE) at equilibrium, the average number $\langle N_{j} \rangle$ of chemical species $j$ is related to the covariance $\langle \delta N_{j}\delta N_{\ell}\rangle$ and free energy $\Phi_{\rm G}$ by
\begin{align}
\label{meantovariance}
\beta \langle \delta N_{j}\delta N_{\ell}\rangle = \partial_{\mu_{\ell}}\langle N_{j}\rangle = -\partial^{2}_{\mu_{j}\mu_{\ell}} \Phi_{\rm G}\ .
\end{align}
Equation~\eqref{meantovariance} implies $\md\mu_{j} = \md\langle N_{j} \rangle/\langle \delta N_{j}^{2}\rangle$, so we can write the designed protocol in terms of the mean copy number, rewriting \eqref{lambdaoptdot2} as
\begin{align}
\md_t \langle N_{j}\rangle^{\rm des} = \frac{\langle \delta N_{j}^{2}\rangle \int_{\langle N_{j}\rangle_{\rm i}}^{\langle N_{j}\rangle_{\rm f}}\md\langle N_{j} \rangle\frac{\sqrt{\zeta(\langle N_{j}\rangle)}}{\langle \delta N_{j}^{2}\rangle} }{\Delta t\sqrt{\zeta(\langle N_{j}\rangle)}} \ .
\end{align}
	
For the two-state closed system, the friction~\eqref{closed friction} is
\begin{align}
\zeta(\langle \Nb\rangle^{\rm c}) &= \frac{\left(\langle \Nb\rangle^{\rm c}\right)^2}{\Ntot k}\left(1-\tfrac{\langle \Nb\rangle^{\rm c}}{\Ntot}\right) \\
&= \frac{\left(\langle \Nb\rangle^{\rm c}\right)^2 \langle \Nub\rangle^{\rm c}}{\Ntot^2 k} \ ,
\end{align}
and the designed protocol becomes
\begin{align}
\md_t \langle &\Nb\rangle^{\rm c~des} = \frac{2}{\Delta t}\sqrt{1-\frac{\langle \Nb\rangle^{\rm c}}{\Ntot}} \\
&\quad \quad \quad \quad \times \left(\sqrt{\Ntot-\langle \Nb\rangle^{\rm c}_{\rm i}}-\sqrt{\Ntot-\langle \Nb\rangle^{\rm c}_{\rm f}}\right) \nonumber \\
&= \frac{2}{\Delta t}\sqrt{\frac{\langle \Nub\rangle^{\rm c}}{\Ntot}} \left(\sqrt{\langle \Nub\rangle^{\rm c}_{\rm i}}-\sqrt{\langle \Nub\rangle^{\rm c}_{\rm f}}\right) \ .
\end{align}

For the two-state open system, the friction~\eqref{openfriction} can be written as 
\begin{align}
\zeta(\langle{N}_{\rm B}\rangle^{\rm o}) = \frac{
\left(\langle{N}_{\rm B}\rangle^{\rm o}\right)^{2}}{\Nbath k} \ ,
\end{align}
which produces a designed protocol for mean copy number:
\begin{align}
\md_t\langle \Nb\rangle^{\rm o~des} = \frac{\Delta \langle \Nb\rangle^{\rm o}}{\Delta t} \ ,
\end{align}
with $\Delta \langle \Nb\rangle^{\rm o}\equiv \langle \Nb\rangle^{\rm o}_{\rm f} -\langle \Nb\rangle^{\rm o}_{\rm i}$. This is equivalent to the naive mean-copy-number protocol.

\section{\label{initial and final velocity}Initial and final protocol velocities} 
Substituting $\mu = \muI$ and $\mu = \muF$ into the designed protocol for the two-state closed system~\eqref{closed lambdaoptdot3} gives the respective initial and final velocities:
\begin{subequations}
\begin{align}
\md_t \mu^{\rm des}|_{\muI} &=\frac{ 2(1+e^{-\muI})}{\Delta t }\left(1-\sqrt{\frac{1+e^{-\muI}}{e^{\Delta\mu}+e^{-\muI}}}\right) \\
\md_t \mu^{\rm des}|_{\muF} &= \frac{ 2(1+e^{-\muF})}{\Delta t }\left(\sqrt{
\frac{e^{\Delta\mu}+e^{-\muI}}{1+e^{-\muI}}}-1\right) \ .	
\end{align}
\end{subequations}

For significant increases in chemical potential ($e^{\muF} \gg e^{\muI}$ and $e^{\muF} \gg 1$), the initial velocity reduces to $\md_t \mu^{\rm des}|_{\muI} \approx 2(1+e^{-\muI})/\Delta t$ and the final to $\md_t \mu^{\rm des}|_{\muF} \approx 2\sqrt{e^{\muF}/(1+e^{\muI})}/\Delta t$. In this limit, the final velocity is exponentially faster in $\muF$ than the initial, because for large chemical potentials the friction is exponentially damped. The opposite limit (large chemical potential decreases) also produces initial velocity independent of $\muF$ and final velocity exponential in $\muF$. 
 
For small changes in chemical potential, Taylor expanding about $\Delta\mu = 0$ gives
\begin{subequations}
\begin{align}
(1+e^{-\muI})\left(1-\sqrt{\frac{1+e^{-\muI}}{e^{\Delta\mu}+e^{-\muI}}}\right) &\approx \Delta\mu/2 \\
(1+e^{-\muF})\left(\sqrt{\frac{e^{\Delta\mu}+e^{-\muI}}{1+e^{-\muI}}}-1\right) &\approx \Delta\mu/2 \ ,
\end{align}
\end{subequations}
so $\md_t \mu^{\rm des}|_{\muI} \approx \md_t \mu^{\rm des}|_{\muF} \approx \Delta\mu/\Delta t$. For sufficiently small changes in chemical potential, the designed protocol reduces to the naive.
	
Comparing the initial and final velocities of the open system~\eqref{openfriction},
\begin{subequations}
\begin{align}
\label{initial open muoptdot}
&\md_t \mu^{\rm des}|_{\muI} = \frac{1}{\Delta t}\left(e^{\Delta\mu}-1\right) \\ \label{final open muoptdot}
&\md_t \mu^{\rm des}|_{\muF} = \frac{1}{\Delta t}\left(1-e^{-\Delta\mu}\right) \ ,
\end{align}	
\end{subequations}
shows that for large chemical potential changes ($\Delta\mu \gg 1$), $\md_t \mu^{\rm des}|_{\muI} \approx e^{\Delta\mu}/\Delta t$ and $\md_t \mu^{\rm des}|_{\muF} \approx 1/\Delta t$, i.e., the initial velocity is exponentially fast, whereas the final velocity is independent of protocol distance. Conversely, for small chemical potential changes, $e^{\Delta\mu}-1\approx \Delta\mu$, and hence $\md_t \mu^{\rm des}|_{\muI} = \md_t \mu^{\rm des}|_{\muF} = \Delta\mu/\Delta t$, reducing to the naive constant-velocity protocol. Therefore, for large $\Delta\mu$ there is an exponential difference in final and initial velocities, whereas for small $\Delta\mu$ there is no difference.

\section{\label{small chemical potential}Work ratio for small chemical potential changes}

For small changes in chemical potential (to lowest order in $\Delta\mu$), the naive excess work~\eqref{Free Energy Work} for a single control parameter is
\begin{subequations}
\begin{align}
W_{\rm ex}^{\rm naive} &= \beta\int_0^{\Delta t}\md t \, \zeta(\mu(t)) \left(\frac{\Delta\mu}{\Delta t}\right)^{2} \nonumber \\
&= \beta\left(\frac{\Delta\mu}{\Delta t}\right)^{2} \int_0^{\Delta t}\md t \\
&\quad \times \left[\zeta(\muI) +\partial_{\mu}\zeta|_{\muI} (\mu(t)-\muI) + \ldots \right] \nonumber \\ 
&= \beta\left(\frac{\Delta\mu}{\Delta t}\right)^{2} \\
&\quad \times \left[\zeta(\muI)\Delta t + \frac{\Delta t}{\Delta\mu}\partial_{\mu}\zeta|_{\muI}\int_{\muI}^{\muF}\md \mu (\mu-\muI) + \ldots \right] \nonumber \\ 
&= \beta\left(\frac{\Delta\mu}{\Delta t}\right)^{2} \\
&\quad \times \left[\zeta(\muI)\Delta t + \frac{\Delta t}{\Delta\mu}\partial_{\mu}\zeta|_{\muI}\frac{1}{2}(\mu-\muI)^2\Big|_{\muI}^{\muF} + \ldots \right] \nonumber \\ 
&= \beta\left(\frac{\Delta\mu}{\Delta t}\right)^{2} \\
&\quad \times \left[\zeta(\muI)\Delta t + \frac{\Delta t}{2\Delta\mu}\partial_{\mu}\zeta|_{\muI}(\Delta\mu)^{2} + \ldots \right] \nonumber \\
&\approx \beta\frac{(\Delta\mu )^{2}}{\Delta t}\zeta(\muI)+ \mathcal{O}\left((\Delta\mu)^3\right) \ ,
\end{align}
\end{subequations}
where the third line follows since the first term is independent of $t$ and the second term is integrated using $\md_t \mu^{\rm naive} = \Delta\mu/\Delta t$ for the naive protocol.

Since the excess work ratio is unity at $\Delta\mu = 0$ and can never decrease below unity, $\Delta\mu = 0$ must be a minimum. Taylor expanding about this minimum gives
\begin{align}
\frac{W_{\rm ex}^{\rm naive}}{W_{\rm ex}^{\rm des}} \equiv R(\Delta\mu) &= 1 + \frac{1}{2} \partial_{\Delta\mu}^{2}R(\Delta\mu)|_{0}(\Delta\mu)^{2} + \mathcal{O}\left((\Delta\mu)^3\right) \ .
\end{align}

\bibliographystyle{apsrev4-1}
%

\begin{thebibliography}{46}%
\makeatletter
\providecommand \@ifxundefined [1]{%
 \@ifx{#1\undefined}
}%
\providecommand \@ifnum [1]{%
 \ifnum #1\expandafter \@firstoftwo
 \else \expandafter \@secondoftwo
 \fi
}%
\providecommand \@ifx [1]{%
 \ifx #1\expandafter \@firstoftwo
 \else \expandafter \@secondoftwo
 \fi
}%
\providecommand \natexlab [1]{#1}%
\providecommand \enquote  [1]{``#1''}%
\providecommand \bibnamefont  [1]{#1}%
\providecommand \bibfnamefont [1]{#1}%
\providecommand \citenamefont [1]{#1}%
\providecommand \href@noop [0]{\@secondoftwo}%
\providecommand \href [0]{\begingroup \@sanitize@url \@href}%
\providecommand \@href[1]{\@@startlink{#1}\@@href}%
\providecommand \@@href[1]{\endgroup#1\@@endlink}%
\providecommand \@sanitize@url [0]{\catcode `\\12\catcode `\$12\catcode
  `\&12\catcode `\#12\catcode `\^12\catcode `\_12\catcode `\%12\relax}%
\providecommand \@@startlink[1]{}%
\providecommand \@@endlink[0]{}%
\providecommand \url  [0]{\begingroup\@sanitize@url \@url }%
\providecommand \@url [1]{\endgroup\@href {#1}{\urlprefix }}%
\providecommand \urlprefix  [0]{URL }%
\providecommand \Eprint [0]{\href }%
\providecommand \doibase [0]{http://dx.doi.org/}%
\providecommand \selectlanguage [0]{\@gobble}%
\providecommand \bibinfo  [0]{\@secondoftwo}%
\providecommand \bibfield  [0]{\@secondoftwo}%
\providecommand \translation [1]{[#1]}%
\providecommand \BibitemOpen [0]{}%
\providecommand \bibitemStop [0]{}%
\providecommand \bibitemNoStop [0]{.\EOS\space}%
\providecommand \EOS [0]{\spacefactor3000\relax}%
\providecommand \BibitemShut  [1]{\csname bibitem#1\endcsname}%
\let\auto@bib@innerbib\@empty
\bibitem [{\citenamefont {Lim}\ \emph {et~al.}(2015)\citenamefont {Lim},
  \citenamefont {Mayer},\ and\ \citenamefont {Pawson}}]{signalling}%
  \BibitemOpen
  \bibfield  {author} {\bibinfo {author} {\bibfnamefont {W.}~\bibnamefont
  {Lim}}, \bibinfo {author} {\bibfnamefont {B.}~\bibnamefont {Mayer}}, \ and\
  \bibinfo {author} {\bibfnamefont {T.}~\bibnamefont {Pawson}},\ }\href@noop {}
  {\emph {\bibinfo {title} {Cell Signaling: Principles and Mechanisms}}}\
  (\bibinfo  {publisher} {Garland Science, Taylor and Francis Group, New
  York},\ \bibinfo {year} {2015})\BibitemShut {NoStop}%
\bibitem [{\citenamefont {Alberts}\ \emph {et~al.}(2002)\citenamefont
  {Alberts}, \citenamefont {Johnson}, \citenamefont {Lewis}, \citenamefont
  {Raff}, \citenamefont {Roberts},\ and\ \citenamefont {Walter}}]{MBOC}%
  \BibitemOpen
  \bibfield  {author} {\bibinfo {author} {\bibfnamefont {B.}~\bibnamefont
  {Alberts}}, \bibinfo {author} {\bibfnamefont {A.}~\bibnamefont {Johnson}},
  \bibinfo {author} {\bibfnamefont {J.}~\bibnamefont {Lewis}}, \bibinfo
  {author} {\bibfnamefont {M.}~\bibnamefont {Raff}}, \bibinfo {author}
  {\bibfnamefont {K.}~\bibnamefont {Roberts}}, \ and\ \bibinfo {author}
  {\bibfnamefont {P.}~\bibnamefont {Walter}},\ }\href@noop {} {\emph {\bibinfo
  {title} {Molecular Biology of the Cell}}}\ (\bibinfo  {publisher} {Garland
  Science, New York},\ \bibinfo {year} {2002})\BibitemShut {NoStop}%
\bibitem [{\citenamefont {Nobelprize.org}(2012)}]{nobel2012}%
  \BibitemOpen
  \bibfield  {author} {\bibinfo {author} {\bibnamefont {Nobelprize.org}},\
  }\href@noop {} {\bibfield  {journal} {\bibinfo  {journal} {The {N}obel
  {P}rize in Chemistry 2012 --- {A}dvanced Information}\ } (\bibinfo {year}
  {2012})}\BibitemShut {NoStop}%
\bibitem [{\citenamefont {Takeda}\ \emph {et~al.}(2002)\citenamefont {Takeda},
  \citenamefont {Kadowaki}, \citenamefont {Haga}, \citenamefont {Takaesu},\
  and\ \citenamefont {Mitaku}}]{nobel1}%
  \BibitemOpen
  \bibfield  {author} {\bibinfo {author} {\bibfnamefont {S.}~\bibnamefont
  {Takeda}}, \bibinfo {author} {\bibfnamefont {S.}~\bibnamefont {Kadowaki}},
  \bibinfo {author} {\bibfnamefont {T.}~\bibnamefont {Haga}}, \bibinfo {author}
  {\bibfnamefont {H.}~\bibnamefont {Takaesu}}, \ and\ \bibinfo {author}
  {\bibfnamefont {S.}~\bibnamefont {Mitaku}},\ }\href {\doibase
  10.1016/S0014-5793(02)02775-8} {\bibfield  {journal} {\bibinfo  {journal}
  {FEBS Lett.}\ }\textbf {\bibinfo {volume} {520}},\ \bibinfo {pages} {97}
  (\bibinfo {year} {2002})}\BibitemShut {NoStop}%
\bibitem [{\citenamefont {Fredriksson}\ \emph {et~al.}(2003)\citenamefont
  {Fredriksson}, \citenamefont {Lagerstr{\"o}m}, \citenamefont {Lundin},\ and\
  \citenamefont {Schi{\"o}th}}]{nobel2}%
  \BibitemOpen
  \bibfield  {author} {\bibinfo {author} {\bibfnamefont {R.}~\bibnamefont
  {Fredriksson}}, \bibinfo {author} {\bibfnamefont {M.~C.}\ \bibnamefont
  {Lagerstr{\"o}m}}, \bibinfo {author} {\bibfnamefont {L.-G.}\ \bibnamefont
  {Lundin}}, \ and\ \bibinfo {author} {\bibfnamefont {H.~B.}\ \bibnamefont
  {Schi{\"o}th}},\ }\href {\doibase 10.1124/mol.63.6.1256} {\bibfield
  {journal} {\bibinfo  {journal} {Molec. Pharmacol.}\ }\textbf {\bibinfo
  {volume} {63}},\ \bibinfo {pages} {1256} (\bibinfo {year}
  {2003})}\BibitemShut {NoStop}%
\bibitem [{\citenamefont {Hasenstaub}\ \emph {et~al.}(2010)\citenamefont
  {Hasenstaub}, \citenamefont {Otte}, \citenamefont {Callaway},\ and\
  \citenamefont {Sejnowski}}]{Hasenstaub:2010eh}%
  \BibitemOpen
  \bibfield  {author} {\bibinfo {author} {\bibfnamefont {A.}~\bibnamefont
  {Hasenstaub}}, \bibinfo {author} {\bibfnamefont {S.}~\bibnamefont {Otte}},
  \bibinfo {author} {\bibfnamefont {E.}~\bibnamefont {Callaway}}, \ and\
  \bibinfo {author} {\bibfnamefont {T.~J.}\ \bibnamefont {Sejnowski}},\
  }\href@noop {} {\bibfield  {journal} {\bibinfo  {journal} {Proc. Natl. Acad.
  Sci. USA}\ }\textbf {\bibinfo {volume} {107}},\ \bibinfo {pages} {12329}
  (\bibinfo {year} {2010})}\BibitemShut {NoStop}%
\bibitem [{\citenamefont {Brown}\ and\ \citenamefont
  {Sivak}(2017{\natexlab{a}})}]{molecularmachines}%
  \BibitemOpen
  \bibfield  {author} {\bibinfo {author} {\bibfnamefont {A.~I.}\ \bibnamefont
  {Brown}}\ and\ \bibinfo {author} {\bibfnamefont {D.~A.}\ \bibnamefont
  {Sivak}},\ }\href@noop {} {\bibfield  {journal} {\bibinfo  {journal} {Physics
  in Canada}\ }\textbf {\bibinfo {volume} {73}},\ \bibinfo {pages} {61}
  (\bibinfo {year} {2017}{\natexlab{a}})}\BibitemShut {NoStop}%
\bibitem [{\citenamefont {Sivak}\ and\ \citenamefont
  {Crooks}(2012)}]{OptimalPaths}%
  \BibitemOpen
  \bibfield  {author} {\bibinfo {author} {\bibfnamefont {D.~A.}\ \bibnamefont
  {Sivak}}\ and\ \bibinfo {author} {\bibfnamefont {G.~E.}\ \bibnamefont
  {Crooks}},\ }\href@noop {} {\bibfield  {journal} {\bibinfo  {journal} {Phys.
  Rev. Lett.}\ }\textbf {\bibinfo {volume} {108}},\ \bibinfo {pages} {190602}
  (\bibinfo {year} {2012})}\BibitemShut {NoStop}%
\bibitem [{\citenamefont {Sivak}\ and\ \citenamefont
  {Crooks}(2016)}]{Workratio}%
  \BibitemOpen
  \bibfield  {author} {\bibinfo {author} {\bibfnamefont {D.~A.}\ \bibnamefont
  {Sivak}}\ and\ \bibinfo {author} {\bibfnamefont {G.~E.}\ \bibnamefont
  {Crooks}},\ }\href@noop {} {\bibfield  {journal} {\bibinfo  {journal} {Phys.
  Rev. E}\ }\textbf {\bibinfo {volume} {94}},\ \bibinfo {pages} {052106}
  (\bibinfo {year} {2016})}\BibitemShut {NoStop}%
\bibitem [{\citenamefont {Darvey}\ and\ \citenamefont
  {Staff}(1966)}]{Closedfirstorder}%
  \BibitemOpen
  \bibfield  {author} {\bibinfo {author} {\bibfnamefont {I.~G.}\ \bibnamefont
  {Darvey}}\ and\ \bibinfo {author} {\bibfnamefont {P.~J.}\ \bibnamefont
  {Staff}},\ }\href {\doibase 10.1063/1.1726855} {\bibfield  {journal}
  {\bibinfo  {journal} {J. Chem. Phys.}\ }\textbf {\bibinfo {volume} {44}},\
  \bibinfo {pages} {990} (\bibinfo {year} {1966})}\BibitemShut {NoStop}%
\bibitem [{\citenamefont {Gadgil}\ \emph {et~al.}(2005)\citenamefont {Gadgil},
  \citenamefont {Lee},\ and\ \citenamefont {Othmer}}]{Firstorderreactions}%
  \BibitemOpen
  \bibfield  {author} {\bibinfo {author} {\bibfnamefont {C.}~\bibnamefont
  {Gadgil}}, \bibinfo {author} {\bibfnamefont {C.~H.}\ \bibnamefont {Lee}}, \
  and\ \bibinfo {author} {\bibfnamefont {H.~G.}\ \bibnamefont {Othmer}},\
  }\href {\doibase https://doi.org/10.1016/j.bulm.2004.09.009} {\bibfield
  {journal} {\bibinfo  {journal} {Bull. Math. Biol.}\ }\textbf {\bibinfo
  {volume} {67}},\ \bibinfo {pages} {901 } (\bibinfo {year}
  {2005})}\BibitemShut {NoStop}%
\bibitem [{\citenamefont {Schnoerr}\ \emph {et~al.}(2015)\citenamefont
  {Schnoerr}, \citenamefont {Sanguinetti},\ and\ \citenamefont
  {Grima}}]{Momentclosure}%
  \BibitemOpen
  \bibfield  {author} {\bibinfo {author} {\bibfnamefont {D.}~\bibnamefont
  {Schnoerr}}, \bibinfo {author} {\bibfnamefont {G.}~\bibnamefont
  {Sanguinetti}}, \ and\ \bibinfo {author} {\bibfnamefont {R.}~\bibnamefont
  {Grima}},\ }\href@noop {} {\bibfield  {journal} {\bibinfo  {journal} {J.
  Chem. Phys.}\ }\textbf {\bibinfo {volume} {143}},\ \bibinfo {pages} {185101}
  (\bibinfo {year} {2015})}\BibitemShut {NoStop}%
\bibitem [{\citenamefont {Lucero}\ \emph {et~al.}(2019)\citenamefont {Lucero},
  \citenamefont {Mehdizadeh},\ and\ \citenamefont {Sivak}}]{Lucero:2019gd}%
  \BibitemOpen
  \bibfield  {author} {\bibinfo {author} {\bibfnamefont {J.~N.~E.}\
  \bibnamefont {Lucero}}, \bibinfo {author} {\bibfnamefont {A.}~\bibnamefont
  {Mehdizadeh}}, \ and\ \bibinfo {author} {\bibfnamefont {D.~A.}\ \bibnamefont
  {Sivak}},\ }\href@noop {} {\bibfield  {journal} {\bibinfo  {journal} {Phys.
  Rev. E}\ ,\ \bibinfo {pages} {012119}} (\bibinfo {year} {2019})}\BibitemShut
  {NoStop}%
\bibitem [{\citenamefont {Gillespie}(2007)}]{Gillespie}%
  \BibitemOpen
  \bibfield  {author} {\bibinfo {author} {\bibfnamefont {D.~T.}\ \bibnamefont
  {Gillespie}},\ }\href@noop {} {\bibfield  {journal} {\bibinfo  {journal}
  {Annu. Rev. Phys. Chem.}\ }\textbf {\bibinfo {volume} {58}},\ \bibinfo
  {pages} {35} (\bibinfo {year} {2007})}\BibitemShut {NoStop}%
\bibitem [{\citenamefont {Gardiner}(1985)}]{Gardiner}%
  \BibitemOpen
  \bibfield  {author} {\bibinfo {author} {\bibfnamefont {C.}~\bibnamefont
  {Gardiner}},\ }\href@noop {} {\emph {\bibinfo {title} {Handbook of Stochastic
  Methods for Physics, Chemistry and the Natural Sciences}}},\ \bibinfo
  {edition} {2nd}\ ed.\ (\bibinfo  {publisher} {Springer, Berlin},\ \bibinfo
  {year} {1985})\BibitemShut {NoStop}%
\bibitem [{\citenamefont {van Kampen}(2007)}]{vanKampen}%
  \BibitemOpen
  \bibfield  {author} {\bibinfo {author} {\bibfnamefont {N.}~\bibnamefont {van
  Kampen}},\ }\href@noop {} {\emph {\bibinfo {title} {Stochastic Processes in
  Physics and Chemistry}}},\ \bibinfo {edition} {3rd}\ ed.\ (\bibinfo
  {publisher} {North-Holland, Amsterdam},\ \bibinfo {year} {2007})\BibitemShut
  {NoStop}%
\bibitem [{\citenamefont {Moler}\ and\ \citenamefont
  {Van~Loan}(1978)}]{matrixexponential}%
  \BibitemOpen
  \bibfield  {author} {\bibinfo {author} {\bibfnamefont {C.}~\bibnamefont
  {Moler}}\ and\ \bibinfo {author} {\bibfnamefont {C.}~\bibnamefont
  {Van~Loan}},\ }\href {\doibase 10.1137/1020098} {\bibfield  {journal}
  {\bibinfo  {journal} {SIAM Rev.}\ }\textbf {\bibinfo {volume} {20}},\
  \bibinfo {pages} {801} (\bibinfo {year} {1978})}\BibitemShut {NoStop}%
\bibitem [{\citenamefont {Moler}\ and\ \citenamefont
  {Van~Loan}(2003)}]{newmatrixexponential}%
  \BibitemOpen
  \bibfield  {author} {\bibinfo {author} {\bibfnamefont {C.}~\bibnamefont
  {Moler}}\ and\ \bibinfo {author} {\bibfnamefont {C.}~\bibnamefont
  {Van~Loan}},\ }\href {\doibase 10.1137/S00361445024180} {\bibfield  {journal}
  {\bibinfo  {journal} {SIAM Rev.}\ }\textbf {\bibinfo {volume} {45}},\
  \bibinfo {pages} {3} (\bibinfo {year} {2003})}\BibitemShut {NoStop}%
\bibitem [{\citenamefont {Gans}(1960)}]{realeigenvalues}%
  \BibitemOpen
  \bibfield  {author} {\bibinfo {author} {\bibfnamefont {P.~J.}\ \bibnamefont
  {Gans}},\ }\href {\doibase 10.1063/1.1731239} {\bibfield  {journal} {\bibinfo
   {journal} {J. Chem. Phys.}\ }\textbf {\bibinfo {volume} {33}},\ \bibinfo
  {pages} {691} (\bibinfo {year} {1960})}\BibitemShut {NoStop}%
\bibitem [{\citenamefont {Wei}\ and\ \citenamefont
  {D.~Prater}(1962)}]{realeigenvaluestwo}%
  \BibitemOpen
  \bibfield  {author} {\bibinfo {author} {\bibfnamefont {J.}~\bibnamefont
  {Wei}}\ and\ \bibinfo {author} {\bibfnamefont {C.}~\bibnamefont
  {D.~Prater}},\ }\href@noop {} {\bibfield  {journal} {\bibinfo  {journal}
  {Adv. Catalysis}\ }\textbf {\bibinfo {volume} {13}},\ \bibinfo {pages} {203}
  (\bibinfo {year} {1962})}\BibitemShut {NoStop}%
\bibitem [{\citenamefont {Brown}\ and\ \citenamefont
  {Sivak}(2017{\natexlab{b}})}]{brownsivak2017}%
  \BibitemOpen
  \bibfield  {author} {\bibinfo {author} {\bibfnamefont {A.~I.}\ \bibnamefont
  {Brown}}\ and\ \bibinfo {author} {\bibfnamefont {D.~A.}\ \bibnamefont
  {Sivak}},\ }\href {\doibase 10.1073/pnas.1707534114} {\bibfield  {journal}
  {\bibinfo  {journal} {Proc. Natl. Acad. Sci. USA}\ }\textbf {\bibinfo
  {volume} {114}},\ \bibinfo {pages} {11057} (\bibinfo {year}
  {2017}{\natexlab{b}})}\BibitemShut {NoStop}%
\bibitem [{\citenamefont {Brown}\ and\ \citenamefont
  {Sivak}(2018)}]{brownsivak2018}%
  \BibitemOpen
  \bibfield  {author} {\bibinfo {author} {\bibfnamefont {A.~I.}\ \bibnamefont
  {Brown}}\ and\ \bibinfo {author} {\bibfnamefont {D.~A.}\ \bibnamefont
  {Sivak}},\ }\href {\doibase 10.1021/acs.jpcb.7b10621} {\bibfield  {journal}
  {\bibinfo  {journal} {J. Phys. Chem. B}\ }\textbf {\bibinfo {volume} {122}},\
  \bibinfo {pages} {1387} (\bibinfo {year} {2018})}\BibitemShut {NoStop}%
\bibitem [{\citenamefont {Schmiedl}\ and\ \citenamefont
  {Seifert}(2008)}]{split1}%
  \BibitemOpen
  \bibfield  {author} {\bibinfo {author} {\bibfnamefont {T.}~\bibnamefont
  {Schmiedl}}\ and\ \bibinfo {author} {\bibfnamefont {U.}~\bibnamefont
  {Seifert}},\ }\href@noop {} {\bibfield  {journal} {\bibinfo  {journal}
  {Europhys. Lett.}\ }\textbf {\bibinfo {volume} {83}},\ \bibinfo {pages}
  {30005} (\bibinfo {year} {2008})}\BibitemShut {NoStop}%
\bibitem [{\citenamefont {Wagoner}\ and\ \citenamefont {Dill}(2016)}]{split2}%
  \BibitemOpen
  \bibfield  {author} {\bibinfo {author} {\bibfnamefont {J.~A.}\ \bibnamefont
  {Wagoner}}\ and\ \bibinfo {author} {\bibfnamefont {K.~A.}\ \bibnamefont
  {Dill}},\ }\href {\doibase 10.1021/acs.jpcb.6b02776} {\bibfield  {journal}
  {\bibinfo  {journal} {J. Phys. Chem. B}\ }\textbf {\bibinfo {volume} {120}},\
  \bibinfo {pages} {6327} (\bibinfo {year} {2016})}\BibitemShut {NoStop}%
\bibitem [{\citenamefont {Dar}\ \emph {et~al.}(2012)\citenamefont {Dar},
  \citenamefont {Razooky}, \citenamefont {Singh}, \citenamefont {Trimeloni},
  \citenamefont {McCollum}, \citenamefont {Cox}, \citenamefont {Simpson},\ and\
  \citenamefont {Weinberger}}]{Dar:2012ab}%
  \BibitemOpen
  \bibfield  {author} {\bibinfo {author} {\bibfnamefont {R.~D.}\ \bibnamefont
  {Dar}}, \bibinfo {author} {\bibfnamefont {B.~S.}\ \bibnamefont {Razooky}},
  \bibinfo {author} {\bibfnamefont {A.}~\bibnamefont {Singh}}, \bibinfo
  {author} {\bibfnamefont {T.~V.}\ \bibnamefont {Trimeloni}}, \bibinfo {author}
  {\bibfnamefont {J.~M.}\ \bibnamefont {McCollum}}, \bibinfo {author}
  {\bibfnamefont {C.~D.}\ \bibnamefont {Cox}}, \bibinfo {author} {\bibfnamefont
  {M.~L.}\ \bibnamefont {Simpson}}, \ and\ \bibinfo {author} {\bibfnamefont
  {L.~S.}\ \bibnamefont {Weinberger}},\ }\href@noop {} {\bibfield  {journal}
  {\bibinfo  {journal} {Proc. Natl. Acad. Sci. USA}\ }\textbf {\bibinfo
  {volume} {109}},\ \bibinfo {pages} {17454} (\bibinfo {year}
  {2012})}\BibitemShut {NoStop}%
\bibitem [{\citenamefont {Tafoya}\ \emph {et~al.}(2019)\citenamefont {Tafoya},
  \citenamefont {Large}, \citenamefont {Liu}, \citenamefont {Bustamante},\ and\
  \citenamefont {Sivak}}]{tafoya2018using}%
  \BibitemOpen
  \bibfield  {author} {\bibinfo {author} {\bibfnamefont {S.}~\bibnamefont
  {Tafoya}}, \bibinfo {author} {\bibfnamefont {S.}~\bibnamefont {Large}},
  \bibinfo {author} {\bibfnamefont {S.}~\bibnamefont {Liu}}, \bibinfo {author}
  {\bibfnamefont {C.}~\bibnamefont {Bustamante}}, \ and\ \bibinfo {author}
  {\bibfnamefont {D.~A.}\ \bibnamefont {Sivak}},\ }\href@noop {} {\bibfield
  {journal} {\bibinfo  {journal} {Proc. Natl. Acad. Sci. USA}\ }\textbf
  {\bibinfo {volume} {116}},\ \bibinfo {pages} {5920} (\bibinfo {year}
  {2019})}\BibitemShut {NoStop}%
\bibitem [{\citenamefont {Gaspard}(2004)}]{Gaspard2004}%
  \BibitemOpen
  \bibfield  {author} {\bibinfo {author} {\bibfnamefont {P.}~\bibnamefont
  {Gaspard}},\ }\href@noop {} {\bibfield  {journal} {\bibinfo  {journal} {J.
  Chem. Phys.}\ }\textbf {\bibinfo {volume} {120}},\ \bibinfo {pages} {8898}
  (\bibinfo {year} {2004})}\BibitemShut {NoStop}%
\bibitem [{\citenamefont {Andrieux}\ and\ \citenamefont
  {Gaspard}(2004)}]{Andrieux2004}%
  \BibitemOpen
  \bibfield  {author} {\bibinfo {author} {\bibfnamefont {D.}~\bibnamefont
  {Andrieux}}\ and\ \bibinfo {author} {\bibfnamefont {P.}~\bibnamefont
  {Gaspard}},\ }\href@noop {} {\bibfield  {journal} {\bibinfo  {journal} {J.
  Chem. Phys.}\ }\textbf {\bibinfo {volume} {121}},\ \bibinfo {pages} {6167}
  (\bibinfo {year} {2004})}\BibitemShut {NoStop}%
\bibitem [{\citenamefont {Andrieux}\ and\ \citenamefont
  {Gaspard}(2007)}]{Andrieux2007}%
  \BibitemOpen
  \bibfield  {author} {\bibinfo {author} {\bibfnamefont {D.}~\bibnamefont
  {Andrieux}}\ and\ \bibinfo {author} {\bibfnamefont {P.}~\bibnamefont
  {Gaspard}},\ }\href@noop {} {\bibfield  {journal} {\bibinfo  {journal} {J.
  Stat. Phys.}\ }\textbf {\bibinfo {volume} {127}},\ \bibinfo {pages} {107}
  (\bibinfo {year} {2007})}\BibitemShut {NoStop}%
\bibitem [{\citenamefont {Vellela}\ and\ \citenamefont
  {Qian}(2008)}]{Vellela2008}%
  \BibitemOpen
  \bibfield  {author} {\bibinfo {author} {\bibfnamefont {M.}~\bibnamefont
  {Vellela}}\ and\ \bibinfo {author} {\bibfnamefont {H.}~\bibnamefont {Qian}},\
  }\href@noop {} {\bibfield  {journal} {\bibinfo  {journal} {J. R. Soc.,
  Interface}\ }\textbf {\bibinfo {volume} {6}},\ \bibinfo {pages} {925}
  (\bibinfo {year} {2008})}\BibitemShut {NoStop}%
\bibitem [{\citenamefont {Ge}\ and\ \citenamefont {Qian}(2009)}]{Ge2009}%
  \BibitemOpen
  \bibfield  {author} {\bibinfo {author} {\bibfnamefont {H.}~\bibnamefont
  {Ge}}\ and\ \bibinfo {author} {\bibfnamefont {H.}~\bibnamefont {Qian}},\
  }\href@noop {} {\bibfield  {journal} {\bibinfo  {journal} {Phys. Rev. Lett.}\
  }\textbf {\bibinfo {volume} {103}},\ \bibinfo {pages} {148103} (\bibinfo
  {year} {2009})}\BibitemShut {NoStop}%
\bibitem [{\citenamefont {Schmiedl}\ and\ \citenamefont
  {Seifert}(2007)}]{Schmiedl2007}%
  \BibitemOpen
  \bibfield  {author} {\bibinfo {author} {\bibfnamefont {T.}~\bibnamefont
  {Schmiedl}}\ and\ \bibinfo {author} {\bibfnamefont {U.}~\bibnamefont
  {Seifert}},\ }\href@noop {} {\bibfield  {journal} {\bibinfo  {journal} {J.
  Chem. Phys.}\ }\textbf {\bibinfo {volume} {126}},\ \bibinfo {pages} {044101}
  (\bibinfo {year} {2007})}\BibitemShut {NoStop}%
\bibitem [{\citenamefont {Rao}\ and\ \citenamefont {Esposito}(2016)}]{Rao2016}%
  \BibitemOpen
  \bibfield  {author} {\bibinfo {author} {\bibfnamefont {R.}~\bibnamefont
  {Rao}}\ and\ \bibinfo {author} {\bibfnamefont {M.}~\bibnamefont {Esposito}},\
  }\href@noop {} {\bibfield  {journal} {\bibinfo  {journal} {Phys. Rev. X}\
  }\textbf {\bibinfo {volume} {6}},\ \bibinfo {pages} {041064} (\bibinfo {year}
  {2016})}\BibitemShut {NoStop}%
\bibitem [{\citenamefont {Falasco}\ \emph {et~al.}(2018)\citenamefont
  {Falasco}, \citenamefont {Rao},\ and\ \citenamefont
  {Esposito}}]{Falasco2018}%
  \BibitemOpen
  \bibfield  {author} {\bibinfo {author} {\bibfnamefont {G.}~\bibnamefont
  {Falasco}}, \bibinfo {author} {\bibfnamefont {R.}~\bibnamefont {Rao}}, \ and\
  \bibinfo {author} {\bibfnamefont {M.}~\bibnamefont {Esposito}},\ }\href
  {\doibase 10.1103/PhysRevLett.121.108301} {\bibfield  {journal} {\bibinfo
  {journal} {Phys. Rev. Lett.}\ }\textbf {\bibinfo {volume} {121}},\ \bibinfo
  {pages} {108301} (\bibinfo {year} {2018})}\BibitemShut {NoStop}%
\bibitem [{\citenamefont {Rao}\ and\ \citenamefont {Esposito}(2018)}]{Rao2018}%
  \BibitemOpen
  \bibfield  {author} {\bibinfo {author} {\bibfnamefont {R.}~\bibnamefont
  {Rao}}\ and\ \bibinfo {author} {\bibfnamefont {M.}~\bibnamefont {Esposito}},\
  }\href {\doibase 10.1063/1.5042253} {\bibfield  {journal} {\bibinfo
  {journal} {J. Chem. Phys.}\ }\textbf {\bibinfo {volume} {149}},\ \bibinfo
  {pages} {245101} (\bibinfo {year} {2018})}\BibitemShut {NoStop}%
\bibitem [{\citenamefont {Machta}(2015)}]{machta15}%
  \BibitemOpen
  \bibfield  {author} {\bibinfo {author} {\bibfnamefont {B.~B.}\ \bibnamefont
  {Machta}},\ }\href@noop {} {\bibfield  {journal} {\bibinfo  {journal} {Phys.\
  Rev.\ Lett.}\ }\textbf {\bibinfo {volume} {115}},\ \bibinfo {pages} {260603}
  (\bibinfo {year} {2015})}\BibitemShut {NoStop}%
\bibitem [{\citenamefont {Bryant}\ and\ \citenamefont
  {Machta}(2020)}]{BryantMachta}%
  \BibitemOpen
  \bibfield  {author} {\bibinfo {author} {\bibfnamefont {S.~J.}\ \bibnamefont
  {Bryant}}\ and\ \bibinfo {author} {\bibfnamefont {B.~B.}\ \bibnamefont
  {Machta}},\ }\href@noop {} {\bibfield  {journal} {\bibinfo  {journal} {Proc.
  Natl. Acad. Sci. USA}\ }\textbf {\bibinfo {volume} {117}},\ \bibinfo {pages}
  {3478} (\bibinfo {year} {2020})}\BibitemShut {NoStop}%
\bibitem [{\citenamefont {Benovic}\ \emph {et~al.}(1986)\citenamefont
  {Benovic}, \citenamefont {Strasser}, \citenamefont {Caron},\ and\
  \citenamefont {Lefkowitz}}]{nobel80}%
  \BibitemOpen
  \bibfield  {author} {\bibinfo {author} {\bibfnamefont {J.~L.}\ \bibnamefont
  {Benovic}}, \bibinfo {author} {\bibfnamefont {R.~H.}\ \bibnamefont
  {Strasser}}, \bibinfo {author} {\bibfnamefont {M.~G.}\ \bibnamefont {Caron}},
  \ and\ \bibinfo {author} {\bibfnamefont {R.~J.}\ \bibnamefont {Lefkowitz}},\
  }\href@noop {} {\bibfield  {journal} {\bibinfo  {journal} {Proc. Natl. Acad.
  Sci. USA}\ }\textbf {\bibinfo {volume} {83}},\ \bibinfo {pages} {2797}
  (\bibinfo {year} {1986})}\BibitemShut {NoStop}%
\bibitem [{\citenamefont {Weller}\ \emph {et~al.}(1975)\citenamefont {Weller},
  \citenamefont {Virmaux},\ and\ \citenamefont {Mandel}}]{nobel81}%
  \BibitemOpen
  \bibfield  {author} {\bibinfo {author} {\bibfnamefont {M.}~\bibnamefont
  {Weller}}, \bibinfo {author} {\bibfnamefont {N.}~\bibnamefont {Virmaux}}, \
  and\ \bibinfo {author} {\bibfnamefont {P.}~\bibnamefont {Mandel}},\
  }\href@noop {} {\bibfield  {journal} {\bibinfo  {journal} {Proc. Natl. Acad.
  Sci. USA}\ }\textbf {\bibinfo {volume} {72}},\ \bibinfo {pages} {381}
  (\bibinfo {year} {1975})}\BibitemShut {NoStop}%
\bibitem [{\citenamefont {Wilden}\ \emph {et~al.}(1986)\citenamefont {Wilden},
  \citenamefont {Hall},\ and\ \citenamefont {Kuhn}}]{nobel82}%
  \BibitemOpen
  \bibfield  {author} {\bibinfo {author} {\bibfnamefont {U.}~\bibnamefont
  {Wilden}}, \bibinfo {author} {\bibfnamefont {S.~W.}\ \bibnamefont {Hall}}, \
  and\ \bibinfo {author} {\bibfnamefont {H.}~\bibnamefont {Kuhn}},\ }\href@noop
  {} {\bibfield  {journal} {\bibinfo  {journal} {Proc. Natl. Acad. Sci. USA}\
  }\textbf {\bibinfo {volume} {83}},\ \bibinfo {pages} {1174} (\bibinfo {year}
  {1986})}\BibitemShut {NoStop}%
\bibitem [{\citenamefont {Benovic}\ \emph {et~al.}(1987)\citenamefont
  {Benovic}, \citenamefont {Kuhn}, \citenamefont {Weyand}, \citenamefont
  {Codina}, \citenamefont {Caron},\ and\ \citenamefont {Lefkowitz}}]{nobel83}%
  \BibitemOpen
  \bibfield  {author} {\bibinfo {author} {\bibfnamefont {J.~L.}\ \bibnamefont
  {Benovic}}, \bibinfo {author} {\bibfnamefont {H.}~\bibnamefont {Kuhn}},
  \bibinfo {author} {\bibfnamefont {I.}~\bibnamefont {Weyand}}, \bibinfo
  {author} {\bibfnamefont {J.}~\bibnamefont {Codina}}, \bibinfo {author}
  {\bibfnamefont {M.~G.}\ \bibnamefont {Caron}}, \ and\ \bibinfo {author}
  {\bibfnamefont {R.~J.}\ \bibnamefont {Lefkowitz}},\ }\href {\doibase
  10.1073/pnas.84.24.8879} {\bibfield  {journal} {\bibinfo  {journal} {Proc.
  Natl. Acad. Sci. USA}\ }\textbf {\bibinfo {volume} {84}},\ \bibinfo {pages}
  {8879} (\bibinfo {year} {1987})}\BibitemShut {NoStop}%
\bibitem [{\citenamefont {von Zastrow}\ and\ \citenamefont
  {Kobilka}(1994)}]{nobel84}%
  \BibitemOpen
  \bibfield  {author} {\bibinfo {author} {\bibfnamefont {M.}~\bibnamefont {von
  Zastrow}}\ and\ \bibinfo {author} {\bibfnamefont {B.~K.}\ \bibnamefont
  {Kobilka}},\ }\href@noop {} {\bibfield  {journal} {\bibinfo  {journal} {J.
  Biol. Chem.}\ }\textbf {\bibinfo {volume} {269}},\ \bibinfo {pages} {18448}
  (\bibinfo {year} {1994})}\BibitemShut {NoStop}%
\bibitem [{\citenamefont {Hein}\ \emph {et~al.}(1994)\citenamefont {Hein},
  \citenamefont {Ishii}, \citenamefont {Coughlin},\ and\ \citenamefont
  {Kobilka}}]{nobel85}%
  \BibitemOpen
  \bibfield  {author} {\bibinfo {author} {\bibfnamefont {L.}~\bibnamefont
  {Hein}}, \bibinfo {author} {\bibfnamefont {K.}~\bibnamefont {Ishii}},
  \bibinfo {author} {\bibfnamefont {S.~R.}\ \bibnamefont {Coughlin}}, \ and\
  \bibinfo {author} {\bibfnamefont {B.~K.}\ \bibnamefont {Kobilka}},\
  }\href@noop {} {\bibfield  {journal} {\bibinfo  {journal} {J. Biol. Chem.}\
  }\textbf {\bibinfo {volume} {269}},\ \bibinfo {pages} {27719} (\bibinfo
  {year} {1994})}\BibitemShut {NoStop}%
\bibitem [{\citenamefont {Feng}\ \emph {et~al.}(2013)\citenamefont {Feng},
  \citenamefont {Zhang}, \citenamefont {Xu}, \citenamefont {Gauron},
  \citenamefont {Ducos}, \citenamefont {Vriz}, \citenamefont {Volovitch},
  \citenamefont {Jullien}, \citenamefont {Weiss},\ and\ \citenamefont
  {Bensimon}}]{accuracy}%
  \BibitemOpen
  \bibfield  {author} {\bibinfo {author} {\bibfnamefont {Z.}~\bibnamefont
  {Feng}}, \bibinfo {author} {\bibfnamefont {W.}~\bibnamefont {Zhang}},
  \bibinfo {author} {\bibfnamefont {J.}~\bibnamefont {Xu}}, \bibinfo {author}
  {\bibfnamefont {C.}~\bibnamefont {Gauron}}, \bibinfo {author} {\bibfnamefont
  {B.}~\bibnamefont {Ducos}}, \bibinfo {author} {\bibfnamefont
  {S.}~\bibnamefont {Vriz}}, \bibinfo {author} {\bibfnamefont {M.}~\bibnamefont
  {Volovitch}}, \bibinfo {author} {\bibfnamefont {L.}~\bibnamefont {Jullien}},
  \bibinfo {author} {\bibfnamefont {S.}~\bibnamefont {Weiss}}, \ and\ \bibinfo
  {author} {\bibfnamefont {D.}~\bibnamefont {Bensimon}},\ }\href@noop {}
  {\bibfield  {journal} {\bibinfo  {journal} {Rep. Prog. Phys.}\ }\textbf
  {\bibinfo {volume} {76}},\ \bibinfo {pages} {072601} (\bibinfo {year}
  {2013})}\BibitemShut {NoStop}%
\bibitem [{\citenamefont {Verdaasdonk}\ \emph {et~al.}(2014)\citenamefont
  {Verdaasdonk}, \citenamefont {Lawrimore},\ and\ \citenamefont
  {Bloom}}]{QuantFluorMic}%
  \BibitemOpen
  \bibfield  {author} {\bibinfo {author} {\bibfnamefont {J.~S.}\ \bibnamefont
  {Verdaasdonk}}, \bibinfo {author} {\bibfnamefont {J.}~\bibnamefont
  {Lawrimore}}, \ and\ \bibinfo {author} {\bibfnamefont {K.}~\bibnamefont
  {Bloom}},\ }\href@noop {} {\bibfield  {journal} {\bibinfo  {journal} {Meth.
  Cell Biol.}\ }\textbf {\bibinfo {volume} {123}},\ \bibinfo {pages} {347}
  (\bibinfo {year} {2014})}\BibitemShut {NoStop}%
\bibitem [{\citenamefont {Salamon}\ and\ \citenamefont
  {Berry}(1983)}]{Salamon:1983:PhysRevLett}%
  \BibitemOpen
  \bibfield  {author} {\bibinfo {author} {\bibfnamefont {P.}~\bibnamefont
  {Salamon}}\ and\ \bibinfo {author} {\bibfnamefont {R.~S.}\ \bibnamefont
  {Berry}},\ }\href@noop {} {\bibfield  {journal} {\bibinfo  {journal} {Phys.
  Rev. Lett.}\ }\textbf {\bibinfo {volume} {51}},\ \bibinfo {pages} {1127}
  (\bibinfo {year} {1983})}\BibitemShut {NoStop}%
\end{thebibliography}

\end{document}